\documentclass[10pt,twocolumn,
						 superscriptaddress,
							 english,
							 pre,
							 showpacs,
							 floatfix,
							 aps]{revtex4-2}

\usepackage{xcolor} 

\usepackage{graphicx}
\usepackage{ntheorem}
\usepackage{mathrsfs}

\usepackage{hyperref}
\usepackage{comment}
\usepackage{bbm}

\usepackage{amsmath}
\usepackage{amssymb}
\usepackage{longtable}
\usepackage{boldline}
\usepackage{braket}

\usepackage{siunitx}

\usepackage{placeins}

\usepackage{array}

\usepackage{lipsum}

\begin{document}
\title{Tricritical point in the quantum Hamiltonian mean-field model}
\author{Harald Schmid}
\affiliation{\mbox{Dahlem Center for Complex Quantum Systems, Freie Universit\"at Berlin, 14195 Berlin, Germany}}
\author{Johannes Dieplinger}
\affiliation{\mbox{Institute for Theoretical Physics, Universit\"at Regensburg, 93040 Regensburg, Germany}}
\author{Andrea Solfanelli}
\affiliation{\mbox{
SISSA and INFN Sezione di Trieste, Via Bonomea 265, I-34136 Trieste, Italy
}}
\author{Sauro Succi}
\affiliation{\mbox{
Center for Life Nano Science @ La Sapienza, Italian Institute of Technology, 00161 Roma, Italy
}}
\affiliation{\mbox{
Physics Department, Harvard University, Oxford Street 17, Cambridge, USA
}}
\author{Stefano Ruffo}
\affiliation{\mbox{
SISSA and INFN Sezione di Trieste, Via Bonomea 265, I-34136 Trieste, Italy
}}
\affiliation{\mbox{
Istituto dei Sistemi Complessi,
Via Madonna del Piano 10, I-50019 Sesto Fiorentino, Italy
}}

\begin{abstract}
{Engineering long-range interactions in experimental platforms has been achieved with great success
in a large variety of quantum systems in recent years. Inspired by this progress, we propose a
generalization of the classical Hamiltonian mean-field model to fermionic particles. We study the
phase diagram and thermodynamic properties of the model in the canonical ensemble for ferromagnetic interactions as a function
of temperature and hopping. At zero temperature, small charge 
fluctuations drive the many-body
system through a first order quantum phase transition from an ordered to a disordered phase. At higher temperatures, the 
fluctuation-induced phase transition remains first
order initially and switches to second order only at a tricritical point. Our results offer an intriguing
example of tricriticality in a quantum system with long-range couplings, which bears direct
experimental relevance. The analysis is performed by exact diagonalization and mean-field theory.}
\end{abstract}
\maketitle
\date{\today}
\section{Introduction}
\label{sec_introduction}
Systems with long-range interactions have been  the subject of considerable interest in both the classical \cite{Campa2014} and the quantum domain \cite{Defenu2021}. 
Besides their thermodynamic features, long-range interactions of quantum bits represents a highly desirable design goal for a universal quantum computer, in order to operate any non-local gate of the network and speed up quantum error correction \cite{Preskill1998}.
A paradigmatic model that has served as a testing bed for different physical phenomena that appear due to long-range interactions is the Hamiltonian mean-field model (HMF) \cite{Antoni1995}. The classical model is exactly solvable in both the canonical and microcanonical ensemble and shows a second order mean-field phase transition when varying the temperature or the energy \cite{Campa2009}. An interesting and open question is what would be a quantum model that plays a similar role to highlight the main features of quantum long-range interactions. Attempts have been made to include semiclassical effects within the original HMF setting. Chavanis \cite{Chavanis2011fermions} studied the zero temperature limit of a Fermi-like distribution, finding that the homogeneous state gains stability with respect to the classical one through a first order phase transition in the quantum constant $h$. The homogeneous state has been found to be stable also for bosons using a more detailed analysis based on a self-consistent Schroedinger equation \cite{Chavanis2011bosons}. Plestid and collaborators have investigated the effects of quantum fluctuations superposed onto the classical behavior of the HMF model for bosons in a series of papers. In a first paper \cite{Plestid2018} they studied quantum interference effects in the violent relaxation phenomenon that appears in the repulsive HMF model by using a Gross-Pitaevskii equation; in a second paper \cite{Plestid2019} the localized solutions of the Gross-Pitaevskii equation were analyzed in full detail; while in a third paper \cite{Plestid2020} they studied the $O(2)$ symmetry of the model and the associated quantum Goldstone modes. 
\begin{figure*}[t!]
   \begin{minipage}{0.48\textwidth}
         \centering
         \includegraphics[width=\textwidth]{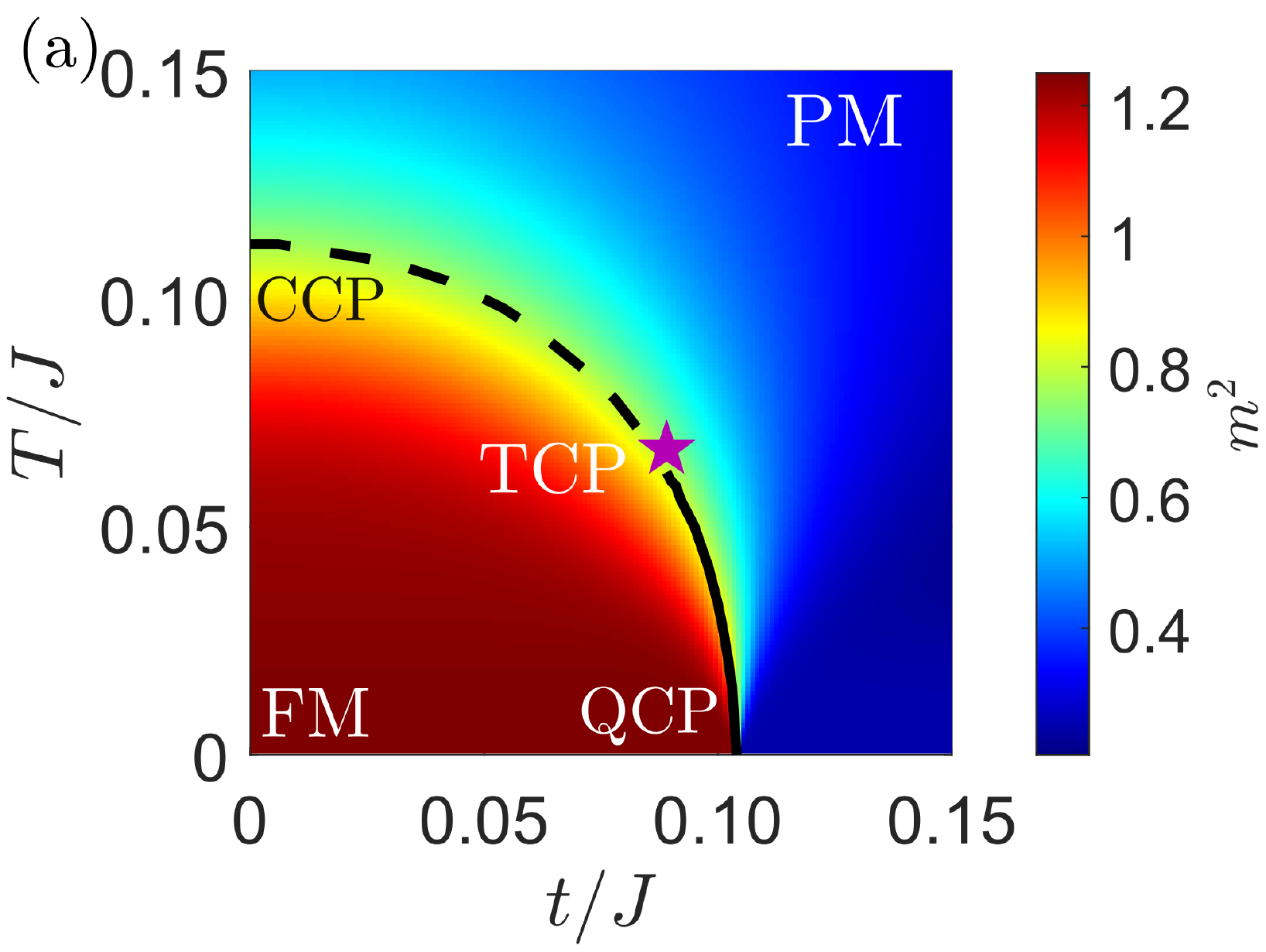}
     \end{minipage}
 	 	\hfill
 	\begin{minipage}{0.48\textwidth}
 		\centering
 		\includegraphics[width=\textwidth]{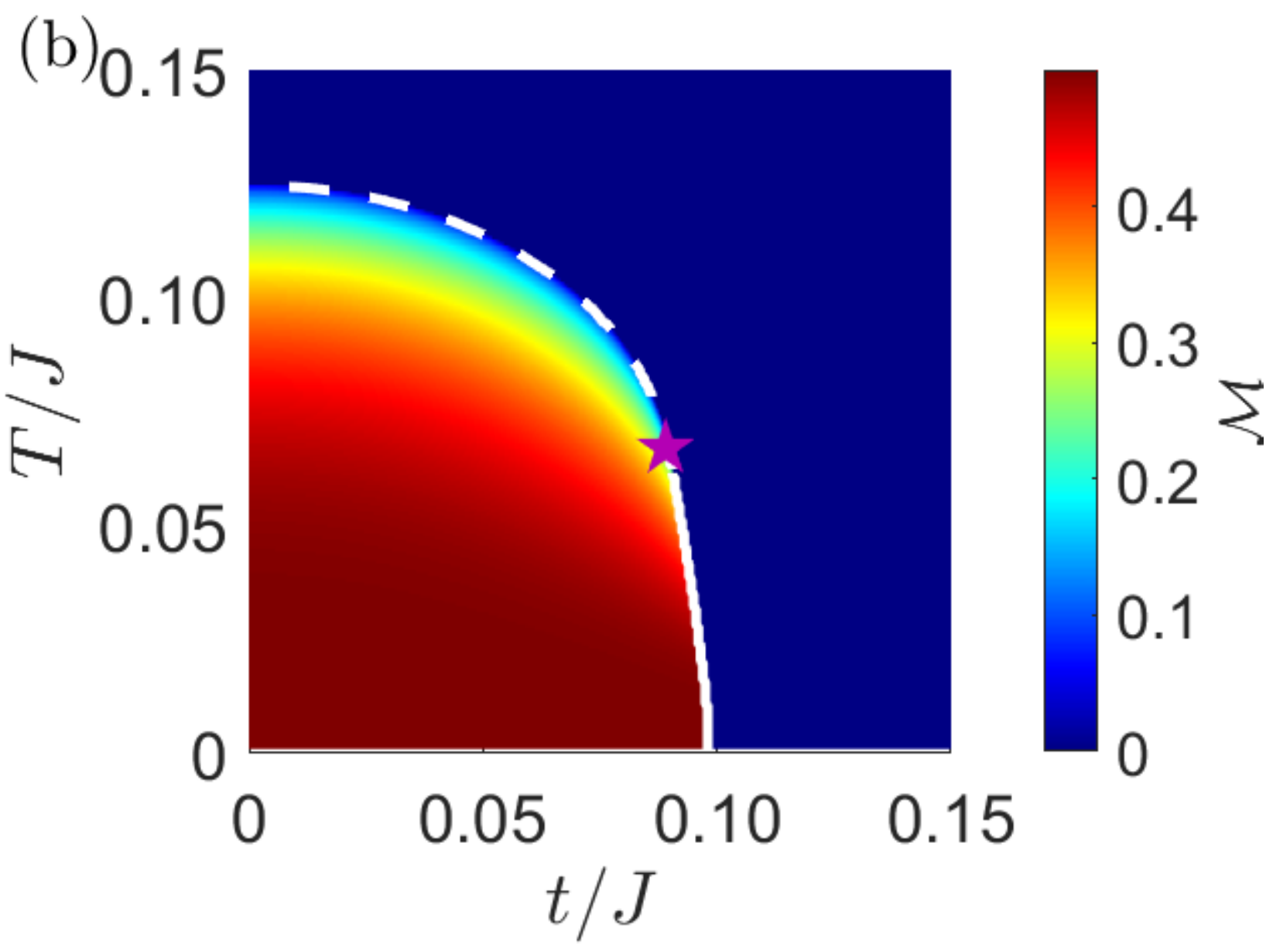}
 	\end{minipage}
         \caption{  Phase diagram of the long-range fermionic Hamiltonian mean-field model at half-filling. (a) Exact diagonalization with $2N=16$ fermions. For small hoppings and small temperatures $T$, the system is ferromagnetically (FM) ordered, for large $t$ and/or $T$ paramagnetically (PM) ordered. A quantum critical point (QCP) is present at zero temperature (horizontal axis) while a classical critical point (CCP) is located on the vertical axis at zero hopping. First order and second order phase boundaries, as indicated by solid and dashed lines respectively, meet at a tricritical point (TCP), marked by a purple star. (b) The phase diagram obtained in mean-field theory shows the order parameter $\mathcal{M}=m/2+\mathcal{O}(1/N)$ (see Eq.~\eqref{eq: order parameter and magnetization}), and is in qualitative agreement with the the finite $N$ exact diagonalization results. The critical points in mean-field theory are:  QCP at $t_c= \pi J/32\approx 0.098 J$, CCP at $T_c= J/8$, TCP at $(t^*,T^*)\approx (0.089,0.067)J$.}
    \label{fig: Phase diagram}
\end{figure*}
\\
On the experimental side quantum long-range interactions can be realized in several forms \cite{Defenu2021}. Some of these experimental settings can be described by models that are closely related to the HMF model, e.g., cold atoms in optical cavities \cite{Baumann2010,Mottl2012,Landig2016}. 
For the classical HMF it was theorized \cite{Schutz2014,Schutz2016} that the model is in experimental reach in a transversely pumped cavity, with atoms ordering into a one-dimensional lattice. 
Their movement around the equilibrium position is essentially semiclassical, described  by an effective Fokker-Planck equation \cite{Domokos2001,Ritsch2013}, and the strong couplings of the atoms to the cavity photons provides effective long-range interactions among them \cite{Schutz2013}. Despite the inherent quantum nature of the long-range interactions, the momentum distribution of the atoms is essentially a classical Maxwell–Boltzmann distribution. Due to the experimental successes for realizing long-range couplings in optical cavities, and the concrete theory of the classical HMF in these systems, we believe that cold atomic systems constitute a promising pathway to realize also a full quantum version of the HMF model. Hereby, it is necessary to find a quantum equivalent of the classical atomic motion, and take the exchange statistics of the particles into consideration.
\\
A second promising experimental architecture, directly related to our proposal, constitutes the recent realization of long-range couplings ($> \SI{1}{\milli\metre}$) between spin qubits fabricated from silicon quantum dots \cite{Mi2018,Borjans2020}.
As for the cold atom experiments, the interactions among qubits is mediated via strong coupling to photons from a microwave resonator via the rules of circuit quantum electrodynamics \cite{Blais2004}. Achieving long-range qubit couplings in the experiment \cite{Borjans2020} relied on the large spin-photon coupling rate, exceeding the cavity decay rate and the spin decoherence rate. Although long-range coupling has been so far achieved only between a single pair of silicon spin qubits (compared to many in cold atomic gases experiments), the fermionic nature of the particles is a priori given which is essential for our model.
\\
Due to these recent experimental results we would like to propose a fully quantum HMF model for fermionic particles. The model we propose represents a strongly interacting many-body system of spin-$\frac{1}{2}$ fermions, where all-to-all $XY$-couplings and charge fluctuations, represented by a conventional hopping term, compete for the ground state.
The system is closely related to the class of $t$-$J$-Hamiltonians \cite{Anderson_RVB,Izyumov_tJmodel} with all-to-all couplings \cite{Kuramoto1991}, with the striking difference that we explicitly permit double occupancy at half-filling to allow for charge fluctuations. We study the model both numerically, by exact diagonalization (ED), using a mean-field approximation and analytical calculations in specific parameter limits. We are able to derive the phase diagram in the plane of the hopping and temperature parameters, showing the presence of a line of quantum phase transitions that are both second and first order and are separated by a tricritical point.
\\
The paper is organized as follows: In Sec.~\ref{sec_model} we introduce the model and draw the connection to the classical HMF model. In Sec.~\ref{sec: phase diagram} we investigate the phase diagram in detail by means of exact diagonalization supported by mean-field theory. Sec.~\ref{sec: mean-field theory} provides a systematic discussion of our mean-field theory treatment. In Sec.~\ref{sec: thermodynamics} we study the thermodynamic properties of the model. Section \ref{sec: Finite size analysis tricritical point} contains a finite size analysis of the numerical data for the tricritical point. We conclude in Sec.~\ref{sec: conclusion}.
%
%
\section{Quantum formulation of the Hamiltonian mean-field model for fermions}
\label{sec_model}

\begin{figure*}[t!]
         \centering
  \includegraphics[width=\textwidth]{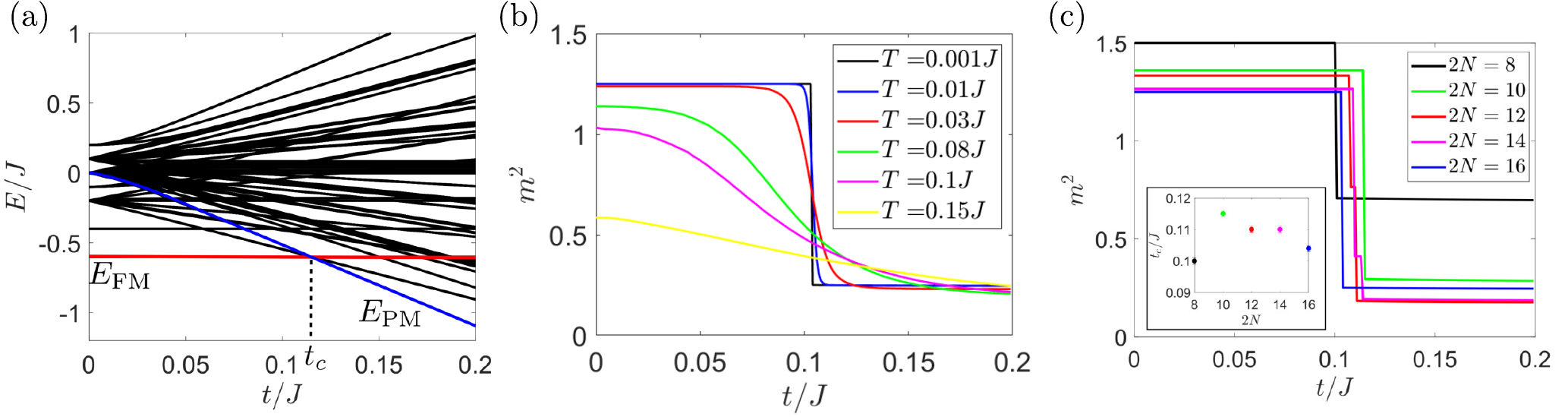}
	\caption{(a) Many-body spectrum at half-filling as a function of hopping for $2N=10$ fermions. The system undergoes a first order quantum phase transition at $t_c$ from a ferromagnetic (FM, red) ground state to a paramagnetic ground state (blue, PM). (b) The magnetization for fixed $T$ displays a clear first order phase transition at zero temperature. (c) Quantum critical point for different system  sizes. For the largest system size $2N=16$ we obtain $t_c/J = 0.104$.}
	\label{fig: spectrum}
\end{figure*}

\subsubsection{Quantum many-body model}
We consider a many-body Hamiltonian with long-range interactions of the form
\begin{align}
\label{eq: quantum hmf}
	&H = H_t + H_{J}\notag\\
	=&   -t\sum\limits^N_{j=1}\sum\limits_{\sigma=\uparrow ,\downarrow} \left(c^\dagger_{j+1,\sigma}c_{j,\sigma}+\mathrm{H.c.}\right)\notag
	  \\
&-\frac{J}{4N}\sum^N_{i<j}\sum_{\alpha,\beta,\gamma,\delta=\uparrow,\downarrow}\left[\sigma^x_{\alpha\beta}\sigma^x_{\gamma\delta}+\sigma^y_{\alpha\beta}\sigma^y_{\gamma\delta}\right]c_{i,\alpha}^\dagger c_{i,\beta}c_{j,\gamma}^\dagger c_{j,\delta}.
\end{align}
The operators $c^\dagger_{j,\sigma}$ ($c_{j,\sigma}$) create (annihilate) particles at site $j$ with spin $\sigma=\uparrow,\downarrow$. By imposing anticommutation relations
$\{c^\dagger_{i,\sigma},c_{j,\sigma'}\} =\delta_{i,j}\delta_{\sigma,\sigma'}$ and $\left\{c_{i,\sigma},c_{j,\sigma'}\right\} =0$, the particles obey fermionic statistics, and periodic boundary conditions are implied. 
The first term in \eqref{eq: quantum hmf} describes nearest-neighbor hopping, and the second term introduces an all-to-all spin-flip interaction. This can be more clearly seen by explicitly inserting the Pauli-matrices $\sigma^a$ ($a=x,y$) in the interaction
\begin{align}
\label{eq: interaction term}
	H_{J}=-\frac{J}{2N}\sum_{i<j}\left(c^\dagger_{i\uparrow}c_{i\downarrow}c^\dagger_{j\downarrow}c_{j\uparrow}+c^\dagger_{i\downarrow}c_{i\uparrow}c^\dagger_{j\uparrow}c_{j\downarrow}\right).
\end{align}
We focus on ferromagnetic (FM) couplings for which a second order phase transition exists in the classical model \cite{Antoni1995}. The $1/N$-factor in the interaction term secures extensivity of the energy. For the most part of the paper, we restrict ourselves to half-filling $\nu=1$, i.e. $N$ particles on $2N$ fermionic sites. Note that in contrast to the well-studied $t$-$J$-model \cite{Anderson_RVB,Izyumov_tJmodel}, we explicitly permit doubly occupied sites. For zero hopping the model becomes equivalent to the  Lipkin-Meshkov-Glick (LMG) model \cite{lipkin1965} at zero field, apart from additional degeneracies due to the fermionic nature of the particles, see App.~\ref{app: zero hopping}. 
\subsubsection{From classical to quantum HMF}
Let us outline the connection of the classical Hamiltonian mean-field model \cite{Antoni1995}
\begin{align}
\label{eq: classic hmf}
    \mathcal{H}=\sum\limits_{i=1}^N\frac{p_i^2}{2}-\frac{J}{2N}\sum\limits_{i,j=1}^N\left[1- \cos(\theta_i-\theta_j)\right],
\end{align} 
with its quantum version. 
In Eq. \eqref{eq: classic hmf} a network of pendula  with individual canonical variables
 $-\pi\leq \theta_i\leq \pi$ and $p_i=\dot{\theta}_i$ interacts in a fully connected way. The analogy to \eqref{eq: quantum hmf} becomes apparent by rewriting the classical potential via two dimensional unit-vectors
\begin{align}
    \mathbf{m}_i=\left(\cos \theta_i ,\sin \theta_i \right).
\end{align}
Using the trigonometric identity $\cos(\theta_i-\theta_j)= \cos(\theta_i)\cos(\theta_j)+\sin(\theta_i)\sin(\theta_j)$ yields a long-range XY spin interaction $V=-(J/2N)\sum_{i,j=1}^N\mathbf{m}_i\cdot \mathbf{m}_j.$ We then identify classical magnetic moments with quantum spin operators in second quantization
$\mathbf{m}_{i,a} \rightarrow \frac{1}{2}
\sum_{\alpha,\beta=\uparrow,\downarrow }c^\dagger_{i\alpha} \sigma^{a}_{\alpha\beta}c_{i\beta}$. The kinetic part in \eqref{eq: classic hmf} is discretized as a nearest neighbor hopping of fermions in \eqref{eq: quantum hmf}. 

At this point, we introduce the magnetization
(density) in the X-Y-plane analogously to Ref. \cite{BotetJullianPRB1983}
\begin{align}
\label{eq: def magnetization}
    m^2 = \frac{1}{(N\mathcal{S})^2}\,\big\langle S_x^2+S_y^2 \big\rangle,
\end{align} 
where  $S_a=\frac{1}{2}\sum_i \sigma^a_{i}$ is the total spin projection in direction $a=x,y,z$ and the prefactor $\mathcal{S}=\frac{1}{2}$ normalizes such that $m^2\leq 1$ in the thermodynamic limit. The magnetization serves as the order parameter. For states with uniform particle density, interaction energy and magnetization are directly linked via 
$4\braket{H_J}=J\left(1-Nm^2/2\right)$. This carries also over to non-uniform densities, see App.\ \ref{app: zero hopping}. The variance of the magnetization 
gives the susceptibility $\chi$ and we measure charge fluctuations with the operator 
\begin{align}
\delta n=\sqrt{\frac{1}{N}\sum_j\braket{\left(\Delta n_j\right)^2}}\,,
\end{align} 
where $\Delta n_j=n_j-\nu$ and $n_j=\sum_{\sigma}c^\dagger_{j,\sigma}c_{j,\sigma}$. Its expectation value gives the local variance in the particle number $\sigma_n^2= \braket{(\delta n)^2}$ at a given filling $\nu$. 
\begin{figure*}[t!]
	\begin{minipage}{0.65\textwidth}
	\includegraphics[width=\textwidth]{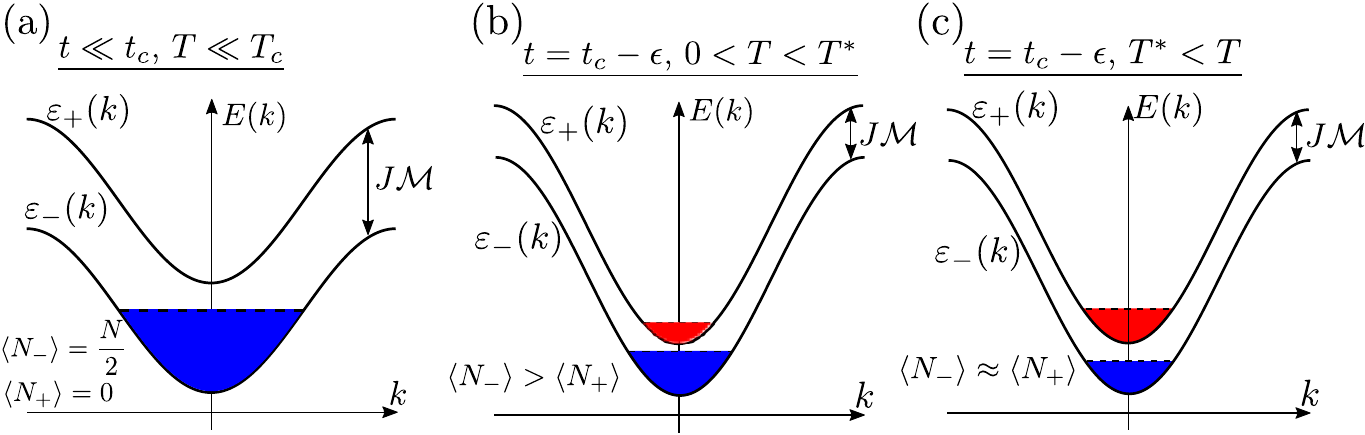}
	\end{minipage}
	\begin{minipage}{0.3\textwidth}
		\centering
		\includegraphics[width=\textwidth]{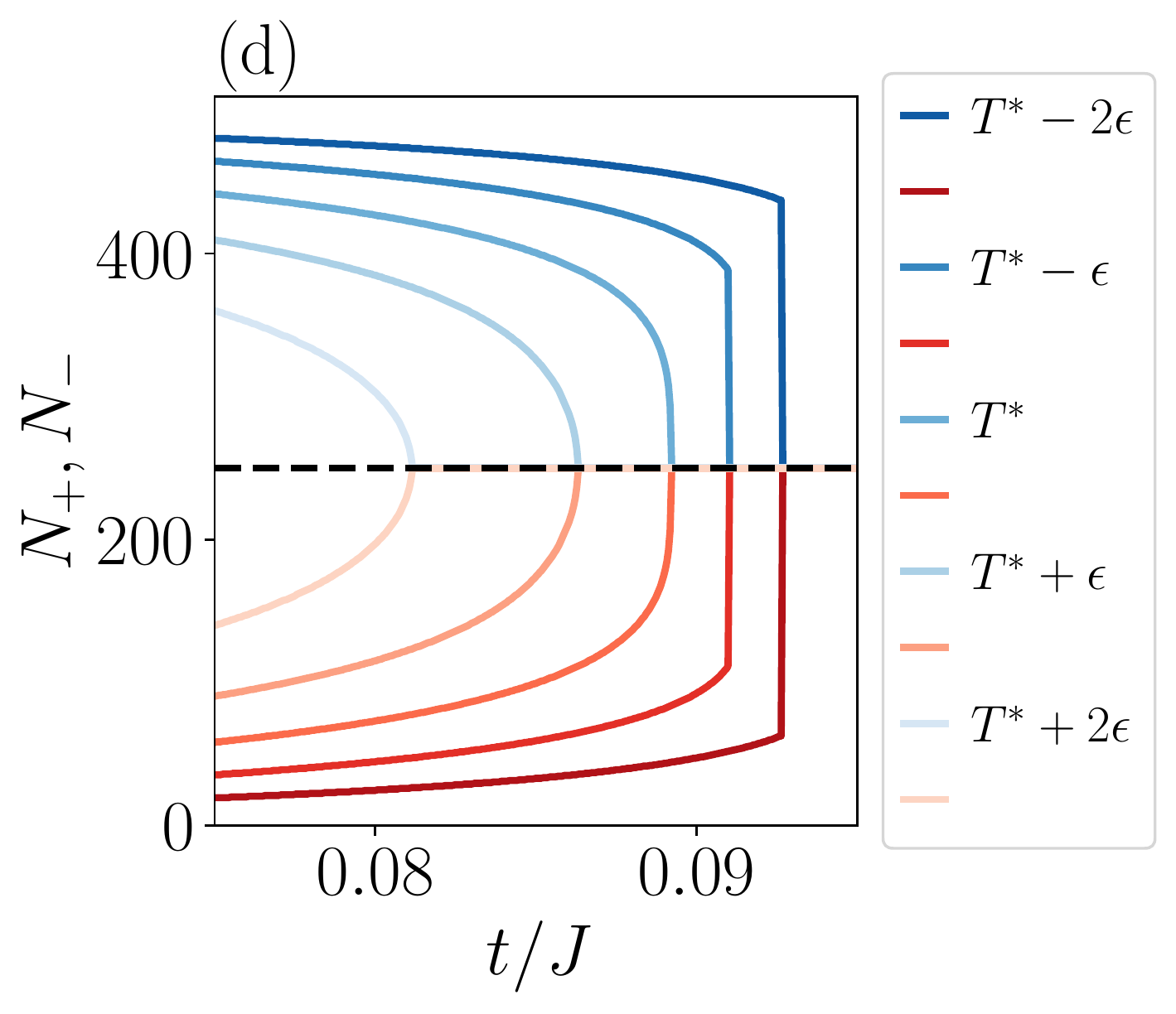}
	\end{minipage}
	\caption{Fillings of the quasi-particle bands in mean-field theory. (a) For $t\ll t_c$ and $T\ll T_c=0$ only the lower is band occupied. (b) Situation for $T<T^*$ in proximity to the phase boundary on the ordered side. The gap has almost closed, but there is a remnant occupation difference (first order phase transition). (c) Situation for $T>T^*$. The gap closes as fast as the occupation equilibrates (second order phase transition). (d) Quasiparticles populations $N_+$ and $N_-$, as calculated in MF theory. The total population is $N = N+a+N_- = 500$ and the black dashed line identifies the transition point when $N_+=N_- = N/2$.}
	\label{fig: populations}
	\label{fig: quasiprticle bands sketch}
\end{figure*}

\subsubsection{Minimal example with $N=2$}
We exemplify the physics of Eq. \eqref{eq: quantum hmf} for a small system with $N=2$ sites, where some characteristics of the infinite lattice model already become visible.  First, we illustrate the quantum nature of the Hamiltonian \eqref{eq: quantum hmf} explicitly.  Let two neighbouring sites be occupied with opposite spin-projections $|\phi\rangle=\ket{\uparrow} \ket{\downarrow}=c^\dagger_{1,\uparrow}c^\dagger_{2,\downarrow}\ket{0}$, and let us denote doubly occupied sites with $|\# \rangle_j \equiv c^\dagger_{j,\uparrow}c^\dagger_{j,\downarrow}|0\rangle$. It is easy to convince oneself that for the specific state that the action of hopping and interaction depends on their relative order, explicitly
\linebreak $H_tH_J|\phi\rangle=-tJ/(2N)[| 0 \rangle| \# \rangle+| \# \rangle| 0 \rangle]$ and  $H_JH_t|\phi\rangle=0$, such that $[H_t,H_J]\neq 0 $. 
Second, we demonstrate that the hopping term reduces magnetic order by simultaneously introducing charge fluctuations. For $t\ll J$, the ground state \linebreak $\ket{\mathrm{FM}}=\frac{1}{\sqrt{2}}\big[\ket{\uparrow}\ket{\downarrow}+\ket{\downarrow}\ket{\uparrow}\big]$ has total spin $S=1$ and signals ferromagnetic order with magnetization density $m^2=2$ at zero temperature. For $J\ll t$, the ground state \linebreak $\ket{\mathrm{PM}}=\frac{1}{2}\big[\ket{\uparrow}\ket{\downarrow}-\ket{\downarrow}\ket{\uparrow}+\ket{\#}\ket{0}+\ket{0}\ket{\#}\big]$ has $S=0$ and is paramagnetically ordered, with lower magnetization $m^2=\frac{1}{2}$. The finite value of of the magnetization for the paramagnetic state reflects the Pauli principle. In constrast, charge fluctuations are more pronounced in the paramagnetic state ($\sigma_n=\frac{1}{4}$) than in the ferromagnetic state ($\sigma_n=0$).

\section{Phase diagram}
\label{sec: phase diagram}

In this section we study the phase diagram  of the quantum HMF, inferred from the magnetization, as a function of temperature and hopping. The coupling $J=1$ serves as the unit of energy.

Fig.~\ref{fig: Phase diagram} (a) shows the magnetization density as defined in Eq. \ref{eq: def magnetization} in the $T$-$t$ plane, obtained by ED for $2N=16$ fermions. 
\footnote{Observables are calculated by fully diagonalizing the Fock-space Hamiltonian matrix \eqref{eq: quantum hmf}. To this end the Hamiltonian is written in occupation number basis of the two spin and $N$ spatial sites and then Jordan-Wigner-transformed to a pure spin Hamiltonian. The Fock-space matrix then has dimension $2^{2N}$, which can be reduced by symmetry considerations. In particular we use particle number conservation, meaning that only blocks of fixed particle number are diagonlized. Throughout the paper, only the half filling block is considered. The implementation is adapted from the code used in Ref. \cite{dieplinger2021} developed for general long-range interacting Hamiltonians. }.
For $t=0$ and $T=0$,  the magnetization is largest and the ground state is ferromagnetically ordered. Increasing the hopping at zero temperature (bottom horizontal), the magnetization stays constant at first, and drops sharply at a critical point $(t,T)=(t_c,0)$, signaling a first order quantum phase transition (QPT), see Fig.~\ref{fig: spectrum} (c). Increasing the temperature at zero hopping (left vertical) reduces the magnetization continuously, and reveals a classical second order phase transition at a critical value $(t,T)=(0,T_c)$ which defines the classical critical point (CCP). The phase boundary emerging from the $T_c$ bends to lower values of the hopping and is found to be second order. Importantly, we find that the phase boundary emerging from $t_c$ at non-zero temperature is first order, and meets in a tricritical point  (TCP) at finite hopping and temperature $(t,T)=(t^*,T^*)$.

We give some insight into the nature of the QPT which occurs due to an abrupt change in the groundstate, defining the quantum critical point (QCP) at $t=t_c$ \cite{Vojta_QPT2003,Sachdev_QPT2011}. In our system the energy of the ferromagnetically ordered state $\ket{\mathrm{FM}}$ matches the energy of a paramagnetically ordered state $\ket{\mathrm{PM}}$, see Fig. \ref{fig: spectrum}. Let us analyze the state  $\ket{\mathrm{FM}}$ in more detail: For $t=0$, the Hamiltonian $H=H_J$ conserves the total spin $S$, the spin-projection $S_z$ and the local variance in particles $\sigma_n$. This gives credit to the fact that the Hamiltonian has block-diagonal sectors in Fock space with equal number of doubly occupied sites. For even $N$, the ground state 
\begin{align}
    |\mathrm{FM}\rangle=\frac{1}{\sqrt{\mathcal{N}}}\left(|\uparrow\downarrow\uparrow\dots >+\mathrm{"all\,transpositions"}\right)
\end{align}
is unique and maximizes the total spin $S=\frac{N}{2}$, while minimizing the spin projection $S_z=0$ and is uniform $\sigma_n=0$. Its energy is
\begin{align}
E_\mathrm{FM}=-\frac{JN}{8}.
\end{align}
and the magnetization is $m^2_\mathrm{FM}= 1+\frac{2}{N}$ ($T=0$ and $t=0$).
A finite finite gap $\delta E = J/(2N)$ \cite{BotetJullianPRB1983} separates it to excited states. Hopping introduces then charge fluctuations by breaking the conservation of local variance in particles, i.e. $\sigma_n$ is no longer a good quantum number. However, because $H_t$ is spin-rotation symmetric and $|\mathrm{FM}\rangle$ is the only state with $S=N/2$ and $S_z=0$, it is completely unaffected by the action of $H_t$ . This is observed in the many-body spectrum shown in Fig. \ref{fig: spectrum} (a). 

%
\begin{figure*}[t!]
		\centering
		\includegraphics[width=\textwidth]{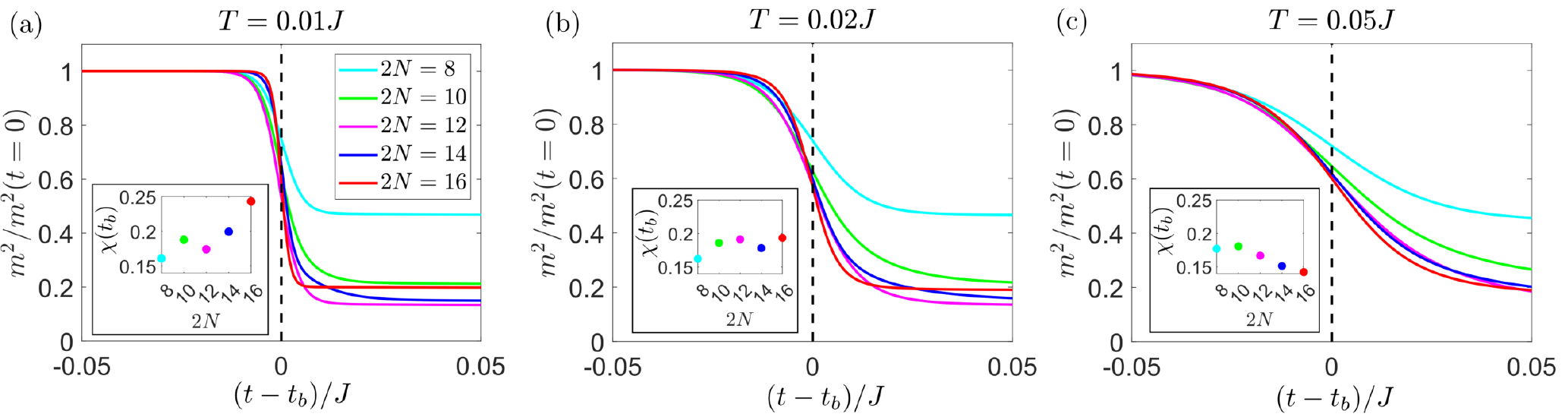}
	\caption{Finite size scaling of the magnetization at the phase boundary  (critical hopping value $t_b$). (a) The slope of the magnetization (susceptibility, inset) grows with $N$ at the phase boundary for fixed and small $T\ll T^*$, signaling a first order phase transition. For larger temperatures the slope saturates (b) with $N$ or even decreases (c).}
	\label{fig: finite size scaling magnetization}
\end{figure*}
%

On the contrary, the paramagnetic state $\ket{\mathrm{PM}}$ with total spin zero is most conveniently treated for $J\ll t$ where it is safely the ground state, see Fig.~\ref{fig: spectrum}. In this limit, the eigenstates of $H\approx H_t=\sum_{k,\sigma}\varepsilon(k) c^\dagger_{k\sigma}c_{k\sigma}$ are approximately plain waves with dispersion is $\varepsilon(k)=-2t\cos(k)$. The paramagnetic state can be constructed by filling up states with momenta $-\frac{\pi}{2}\leq k \leq \frac{\pi}{2}$, and we get by integration in the thermodynamic limit
\begin{align}
E_\mathrm{PM}\simeq -\frac{4Nt}{\pi}.
\end{align}
This result holds up to second order in the coupling $\mathcal{O}\left(\frac{J^2}{tN^2}\right)$ and in the thermodynamic limit, for details see Sec.~\ref{sec: mean-field theory}. By equating the ground state energies, we get an estimate for the critical hopping
\begin{align}
\label{eq: mean-field for tc}
\frac{t_c}{J}\simeq \frac{\pi}{32} \approx 0.0982.
\end{align}
This is reproduced by the mean-field approximation in Sec. \ref{sec: mean-field theory} and agrees well with the numerical result for $2N=16$ fermions with a value $t_c/J=0.104 $. In Fig.~\ref{fig: spectrum} we see a decreasing trend to slightly smaller $t_c$ for larger system sizes in the numerics.
We stress that the quantum phase transition is present for every system size.
\\
Next, we discuss the first order phase transition for finite $T<T^*$  and the TCP. These features of the phase diagram constitute the key findings of this paper, and are best understood within a Hartree mean-field (MF) decoupling of the interaction term.  Here, we present only the main aspects of the MF analysis, for details see Sec.~\ref{sec: mean-field theory}. The MF Hamiltonian reads in momentum space and in the thermodynamic limit
\begin{align}
\label{eq: H_MF}
	H_{\mathrm{MF}} &= \sum_{k}\sum_{\sigma\sigma^\prime}c^\dagger_{k\sigma}h_{\sigma\sigma^\prime}(k)c_{k\sigma^\prime} +\frac{JN\mathcal{M}^2}{2},
	\\
	\notag
	 h_{\sigma\sigma^\prime}(k) &=\varepsilon(k)\delta_{\sigma\sigma^\prime}
	-\frac{J\mathcal{M}}{2}\,\bigg(\sigma^x_{\sigma\sigma^\prime} \cos(\varphi)-\sigma^y_{\sigma\sigma^\prime} \sin(\varphi)\bigg).
\end{align}
We have defined the uniform order parameter $\mathcal{M}= \delta_{ij}\braket{c^\dagger_{i,\uparrow} c_{j,\downarrow}}e^{-i\varphi}$ with real amplitude $\mathcal{M}$ and phase $\varphi$. The phase $\varphi$ can take any value because the direction of the magnetization can be chosen arbitrarily in the $XY$-plane. In fact, fixing $\varphi$ breaks the continuous spin-rotation symmetry, as expected for ferromagnetic order. 
Eq. \eqref{eq: H_MF} is diagonalised by quasiparticles with spin pointing in the direction of the order parameter and dispersion $\varepsilon_\pm (k)=\varepsilon(k)\pm \frac{J\mathcal{M}}{2}$. The two bands are separated by a gap which depends on the order parameter. The self-consistency equation for $\mathcal{M}$ is given by the difference in band occupation   
\begin{align}
	\mathcal{M} &=\frac{\nu_- - \nu_+}{2},
\end{align}
with $\nu_\pm=\frac{1}{N}\braket{\sum_k c^\dagger_{k \pm} c_{k \pm}}$. The band occupation as a function of hopping is shown  in Fig. \ref{fig: populations} for designated values of $T$. At $T=0$, the lower (upper) band is completely filled (empty) for any $t<t_c=\frac{\pi}{32}$. For $t=t_c$, the gap closes and the occupations equalize abruptly, signaling a first order phase transition in agreement with the results from ED. At finite but small $T<T^*$, the lower band is continuously depleted for $t<t_c$ and fills up the upper band. At the critical point, however, there remains a residual difference in occupation such that the drop in the order parameter is still discontinuous. Hence, the phase transition stays first order. The behavior changes at the tricritical point
\begin{align}
(t^*,T^*)\approx (0.089,0.067)J,
\end{align}
which is determined from the free energy, see Sec.~\ref{sec: mean-field theory}.
The difference in occupation vanishes when the phase boundary is reached such that the nature of the phase transition becomes second order. Notice that the first order transition appears in a very fine-tuned region of hopping values $t \in [t^*,t_c]$.
\\
Signatures of a change in the order of the phase transition are also present in the numerical data and are identified by a finite-size scaling. Fig. \ref{fig: finite size scaling magnetization} shows that for small $T<T^*$ the magnetization becomes step-like as a function of $N$ at the phase boundary. Therefore, the susceptibility diverges at $t_c$, in agreement with the mean-field theory prediction of a first order phase transition. For $T<T^*\approx 0.04J$ the slope of the maximum slope of the  magnetization at the phase  phase boundary saturates, implying a second order phase transition. For a more detailed analysis of the numerical data, including an extrapolation to the $N\to \infty$ limit, we refer the reader to Sec.~\ref{sec: Finite size analysis tricritical point}.
\begin{figure*}[t!]
	\begin{minipage}{0.48\textwidth}
		\centering
	\includegraphics[width=\textwidth]{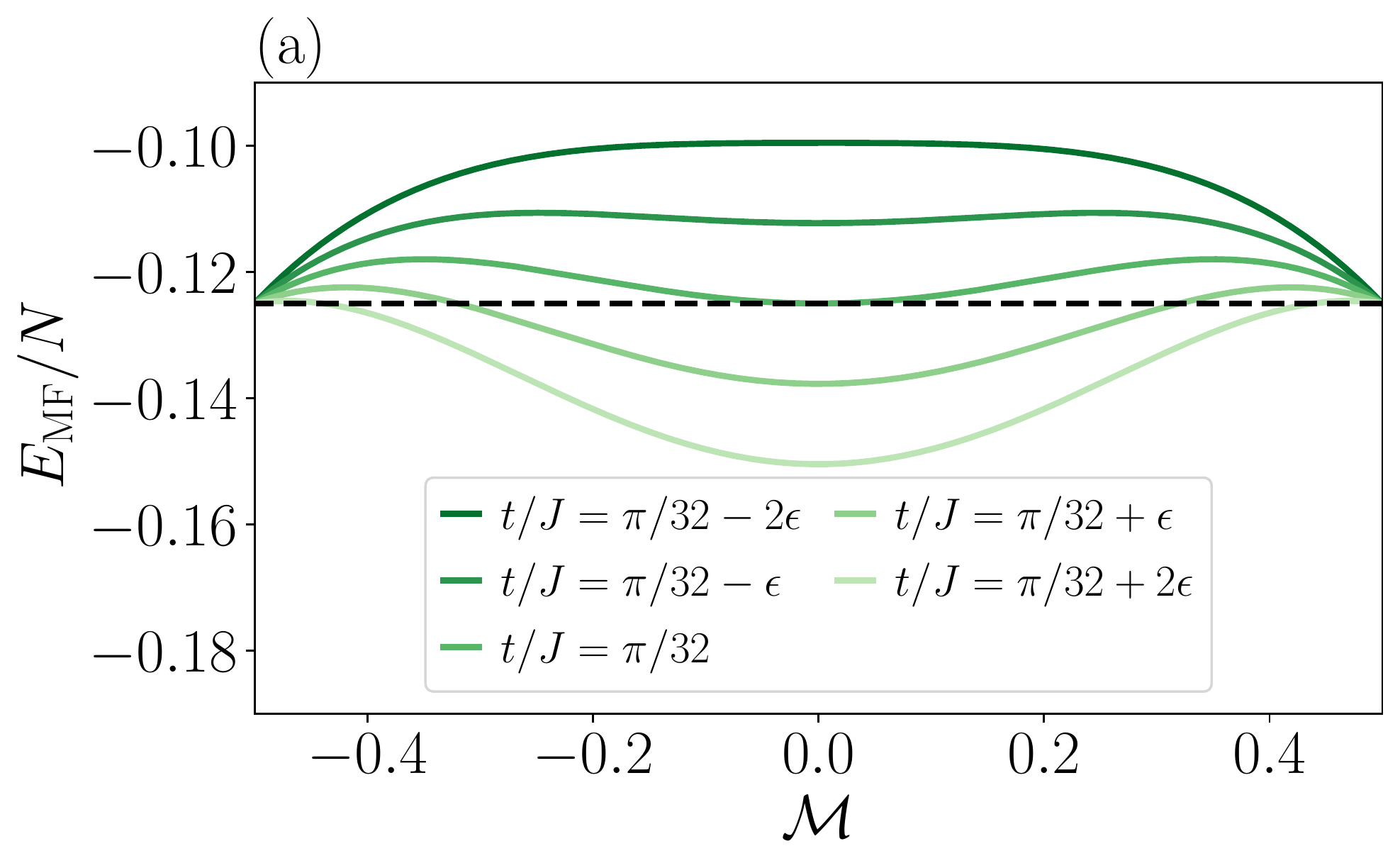}
	\end{minipage}
	\hfill
	\begin{minipage}{0.48\textwidth}
		\includegraphics[width=\textwidth]{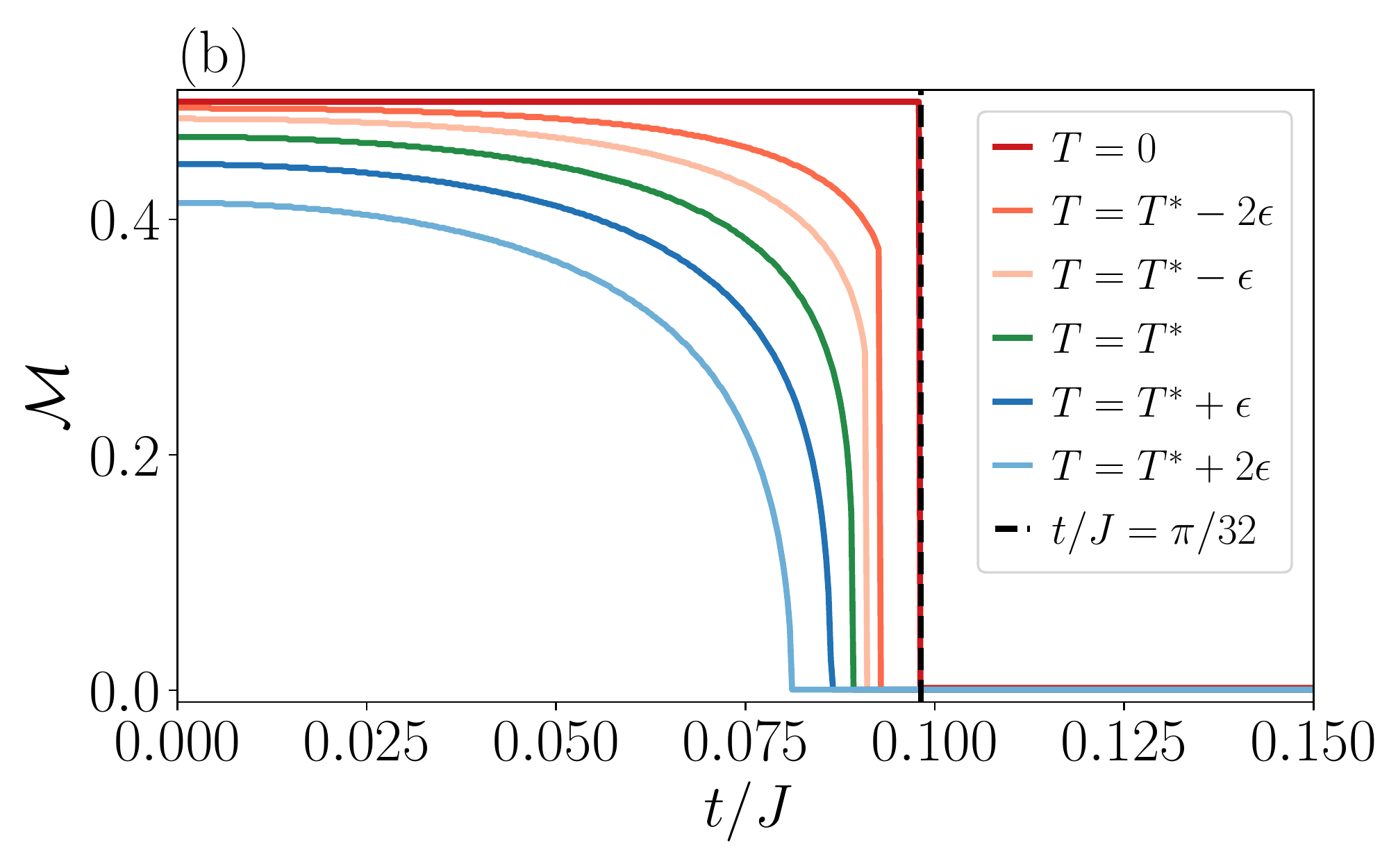}
	\end{minipage}
	\caption{(a) Mean-field energy density $E_{\mathrm{MF}}/N$ as a function of the order parameter $\mathcal{M}$ for different hopping values in the proximity ($\epsilon = 0.01$) of the QCP. (b) Mean-field order parameter $\mathcal{M}$ as a function of the hopping amplitude $t/J$, for different values of the temperature. The black dashed line indicates the QCP.}
	\label{fig: emf}
\end{figure*}
\\
Finally, we briefly discuss the vertical temperature axis of the phase diagram at $t=0$ where a classical second order phase transition occurs at the critical point $T_c$ (CCP). This phase transition is also present in the classical HMF model and is attributed to the long-range nature of the interaction: it is well-known that the short-range XY model has no phase transition for dimensions $D<2$ \cite{Kosterlitz1973}. The transition temperature can be exactly calculated in mean-field theory 
\begin{align}
\frac{T_c}{J}=\frac{1}{8}.
\end{align}
Our numerical data (ED) $T_c/J=0.112\pm 0.001$ agrees for $2N=16$ sites with the MF result within 10\%. This assures us that we are well in the thermodynamic limit. For $t=0$ we have also obtained the spectrum by symmetry considerations and constructed  the degeneracies combinatorically, see App.~\ref{app: zero hopping}. This gives us access to the partition function up to $2N=40$ fermions and serves as another crosscheck to the numerics. 
\section{Mean-field theory}
\label{sec: mean-field theory}
%
%
%
%
In this section we study the mean-field solution of the model. In particular, we consider the interaction term of the quantum many-body Hamiltonian \eqref{eq: interaction term}. 
In the mean-field decoupling scheme we use a Hartree approximation to replace the quartic terms in the interaction according to
\begin{subequations}
	\begin{align}
		c^\dagger_{i,\uparrow}c_{i,\downarrow}c^\dagger_{j,\downarrow}c_{j,\uparrow}&\simeq c^\dagger_{i,\uparrow}c_{i,\downarrow}\langle c^\dagger_{j,\downarrow}c_{j,\uparrow}\rangle+c^\dagger_{j,\downarrow}c_{j,\uparrow}\langle c^\dagger_{i,\uparrow}c_{j,\downarrow}\rangle\notag\\
		&-\langle c^\dagger_{i,\uparrow}c_{i,\downarrow}\rangle\langle c^\dagger_{j,\downarrow}c_{j,\uparrow}\rangle,\\
		c^\dagger_{i,\downarrow}c_{i,\uparrow}c^\dagger_{j,\uparrow}c_{j,\downarrow}&\simeq c^\dagger_{i,\downarrow}c_{i,\uparrow}\langle c^\dagger_{j,\uparrow}c_{j,\downarrow}\rangle+c^\dagger_{j,\uparrow}c_{j,\downarrow}\langle c^\dagger_{i,\downarrow}c_{i,\uparrow}\rangle\notag\\
		&-\langle c^\dagger_{i,\downarrow}c_{i,\uparrow}\rangle\langle c^\dagger_{j,\uparrow}c_{j,\downarrow}\rangle.
	\end{align}
	\label{eq: mean-field_Hartee}
\end{subequations}
Fock terms $\langle c^\dagger_{i,\sigma}c_{j,\sigma}\rangle$ are not taken into account, because they provide only finite size corrections, as shown in App.~\ref{sec: app fock term}. We introduce the order parameter $\mathcal{M}$ defined as
\begin{align}
	\mathcal{M} = \langle c^\dagger_{j,\uparrow}c_{j,\downarrow}\rangle e^{-i\varphi} = \langle c^\dagger_{j,\downarrow}c_{j,\uparrow}\rangle e^{i\varphi}.\label{eq: mean-field m}
\end{align}
%

\begin{figure*}[t!]
		\centering
	\includegraphics[width=\textwidth]{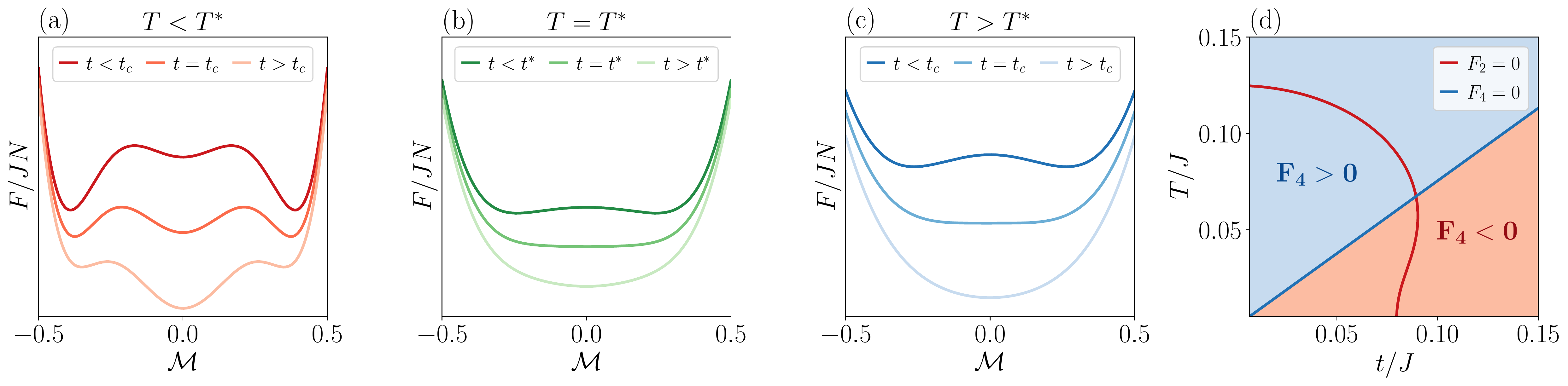}
	\caption{Mean-field free energy $F$ in the proximity of the critical point for different values of temperatures and hoppings. (a) For $T<T^*$, $F$ has three minima which become degenerate at the critical point, signaling a first order phase transition. (b) At $T=T^*$ $F$  displays a flat minimum at the critical point, since both the second and fourth derivative become zero. The transition becomes second order. (c) For $T>T^*$ $F$ passes from a double well shape for to a single well for different hoppings. The transition is second order. (d) Signs of the coefficient $F_4$ are indicated by different shadings. $F_2$ ($F_4$) vanishes on the red (blue) line. Their intersection marks the TCP.}
	\label{fig:Free energy}
\end{figure*}
$\mathcal{M}$ is site independent, because the Hamiltonian is translationally invariant. This is related to the magnetization, introduced in Eq. \ref{eq: def magnetization}, through 
\begin{align}
\label{eq: order parameter and magnetization}
m^2 = 4\mathcal{M}^2(N-1)/N+2/N,
\end{align}
which in the thermodynamic limit gives $\mathcal{M}\simeq m/2$. The mean-field version of the Hamiltonian in Eq.~\eqref{eq: H_MF} is obtained by passing to Fourier space. This Hamiltonian is quadratic and then it can be easily diagonalized  by introducing the fermionic quasiparticle operators 
\begin{align}
	c_{k,\pm} = \frac{c_{k,\uparrow}e^{i\varphi}\pm c_{k,\downarrow}e^{-i\varphi}}{\sqrt{2}}.
	\label{eq: pm quasiparticles}
\end{align}
These are nothing but the fermionic creation and annihilation operators for particles with momentum $k$, whose spin state is an eigenstate of the $\sigma_{xy}(\varphi) = \cos(\varphi)\sigma^x-\sin(\varphi)\sigma^y$ operator, i.e., $c^\dagger_{k,\pm}|0\rangle = |k,s_{xy}(\varphi) = \pm\rangle.$
Leading to the diagonal form of the Hamiltonian 
\begin{align}
	H_{\mathrm{MF}} = \sum_{k,\pm}\varepsilon_{\pm}(k,\mathcal{M})c_{k,\pm}^\dagger c_{k,\pm}+\frac{\mathcal{M}^2J}{2}(N-1),
	\label{eq:diagonal mean field H}
\end{align}
where the quasiparticle spectrum, as previously stated, is made by two cosine bands separated by $\mathcal{M}$   
\begin{align}
	&\varepsilon_{\pm}(k,\mathcal{M}) = -2t\cos(k)\mp \mathcal{M}J\frac{N-1}{2N}.\label{eq: quasiparticles spectrum}
\end{align}
\subsubsection{Zero temperature mean-field theory}
At zero temperature we are interested in the ground state of the system, and we can variationally minimize $E_{\mathrm{MF}} = \langle \psi_{\mathrm{MF}}|H_{\mathrm{MF}}|\psi_{\mathrm{MF}}\rangle$. The condition $\partial E_{\mathrm{MF}}/\partial \mathcal{M}=0$ provides a self-consistent expression for the order parameter
\begin{align}
	\mathcal{M} = \frac{1}{2N}\sum_k\left(n_{k,+}-n_{k,-}\right) = \frac{\nu_+-\nu_-}{2},
	\label{eq:selfconsistent_m}
\end{align}
where $n_{k,\pm} = \langle c_{k,\pm}^\dagger c_{k,\pm} \rangle$, and $\nu_{\pm} =\frac{1}{N}\sum_k n_{k,\pm} $ is the density of $(\pm)$ quasiparticles. Figure \ref{fig: emf} (a) shows the mean-field energy density $E_{\mathrm{MF}}/N$ as a function of the order parameter for different values of $t/J$ around the critical point. We notice that at $t/J = (t/J)_c$ the minimum of $E_{\mathrm{MF}}$ suddenly jumps from $\mathcal{M} = \pm1/2$ to $\mathcal{M} = 0$, thus signaling a first order quantum phase transition. For $N\gg 1$ we can perform a continuum limit in $k$ which allows us to exactly compute $E_{\mathrm{MF}}$ in the two opposite situations
\begin{align}
	&E_{\mathrm{MF}}(\mathcal{M} = 0) = -4Nt\int_{-\frac{\pi}{2}}^{\frac{\pi}{2}}\frac{dk}{2\pi}\cos(k) =  -\frac{4Nt}{\pi},\\
	&E_{\mathrm{MF}}\left(\mathcal{M} = \frac{1}{2}\right) =-4Nt\int_{-\pi}^{\pi}\frac{dk}{2\pi}\cos(k)-\frac{JN}{8} =-\frac{JN}{8}.
\end{align}
The critical point is identified by the condition $E_{\mathrm{MF}}(0) = E_{\mathrm{MF}}(1/2)$ leading to $\left(t/J\right)_c = \pi/32$, as shown in Fig. \ref{fig: emf} (b), where the zero temperature order parameter corresponds to the dark red line. The order parameter $\mathcal{M} = \mathrm{argmin}[E_{\mathrm{MF}}(\mathcal{M})]$ displays a discontinuous jump at this value, corresponding to a first order quantum critical point. We notice that the mean-field results are in good agreement with the numerical analysis presented in the previous section for a finite system.

\subsubsection{Finite temperature mean-field theory}
\begin{figure*}[t!]
            \begin{minipage}{0.32\textwidth}
         \centering
         \includegraphics[width=\textwidth]{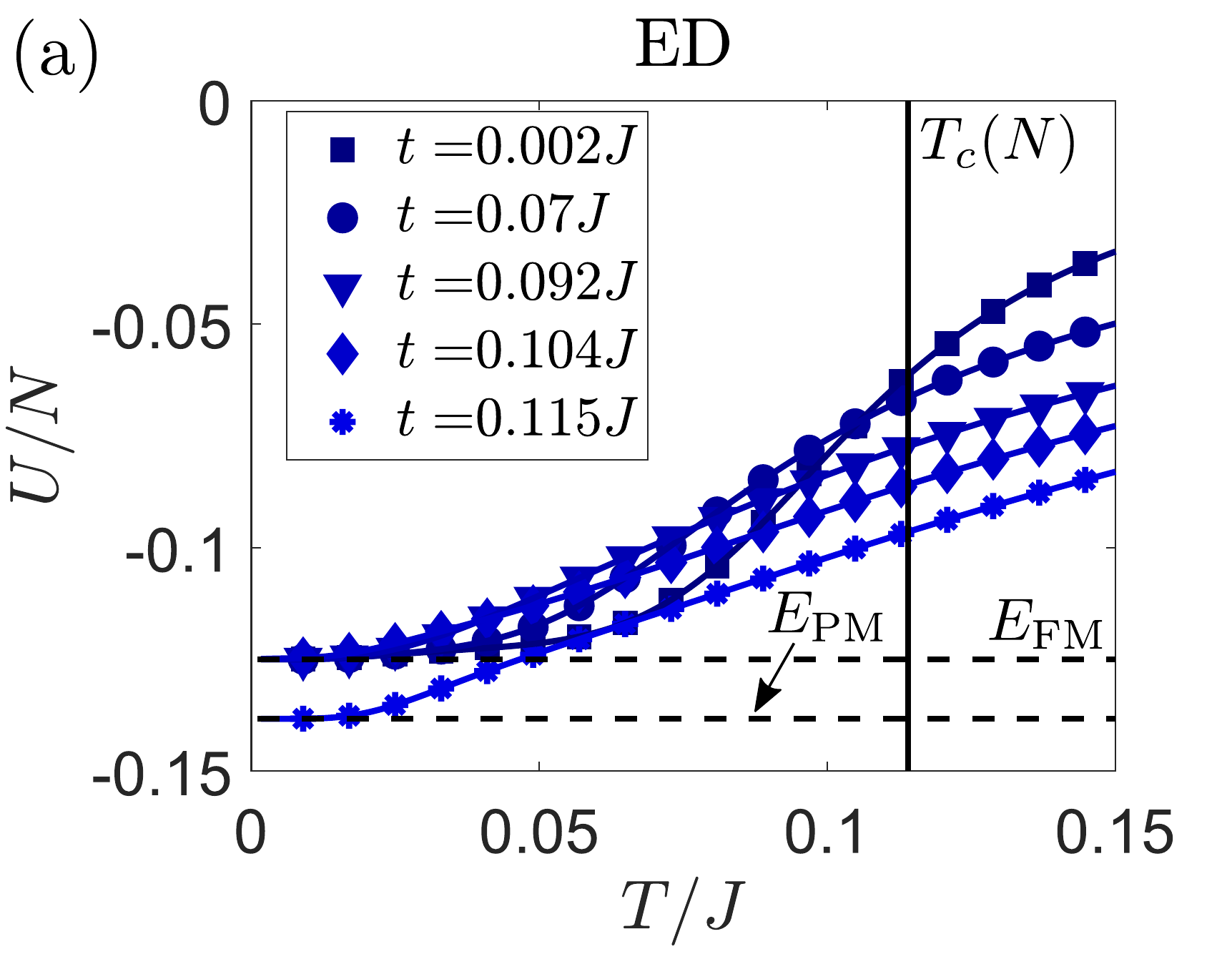}
     \end{minipage}
     \hfill
   \begin{minipage}{0.32\textwidth}
         \centering
         \includegraphics[width=\textwidth]{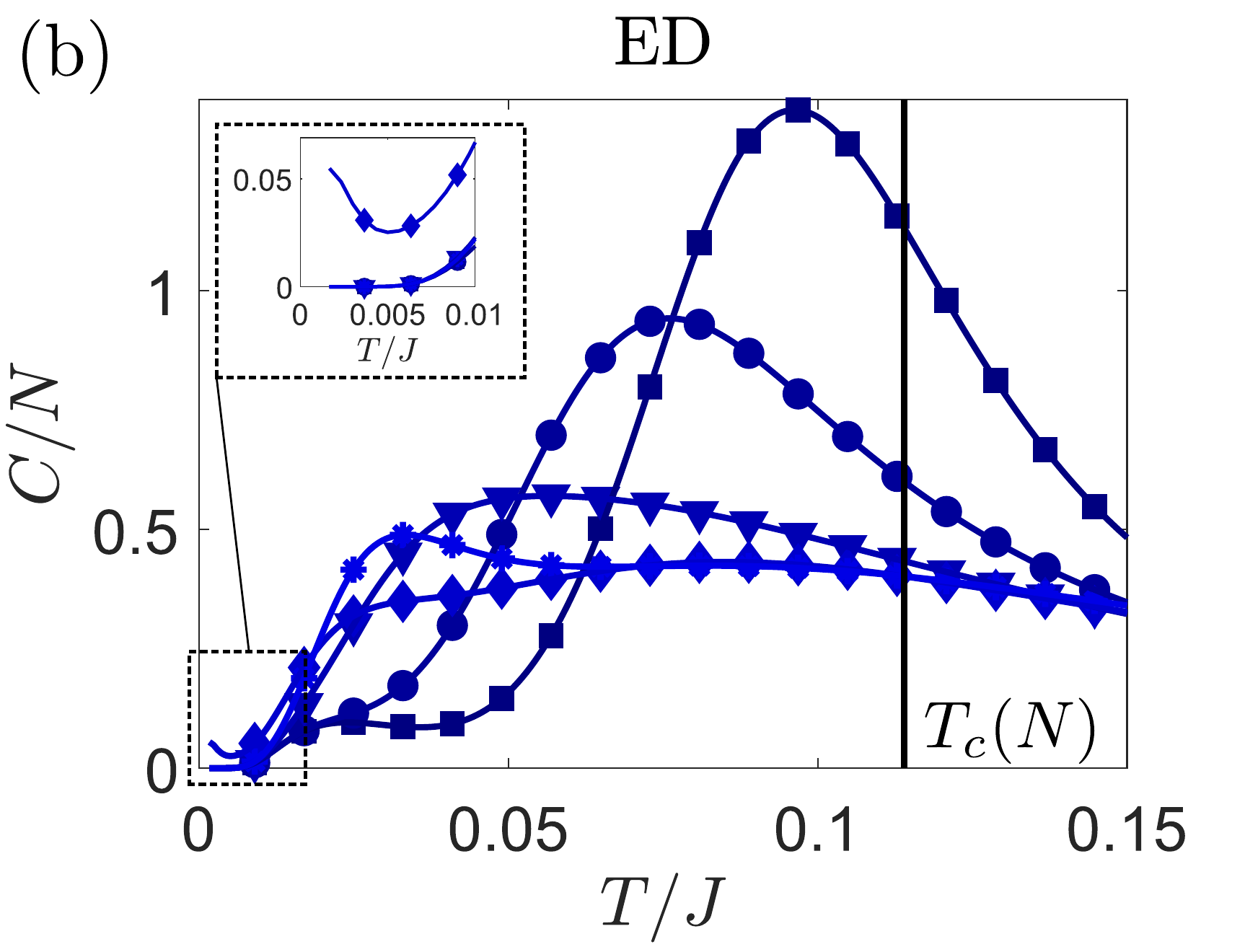}
     \end{minipage}
         \hfill
            \begin{minipage}{0.32\textwidth}
         \centering
         \includegraphics[width=\textwidth]{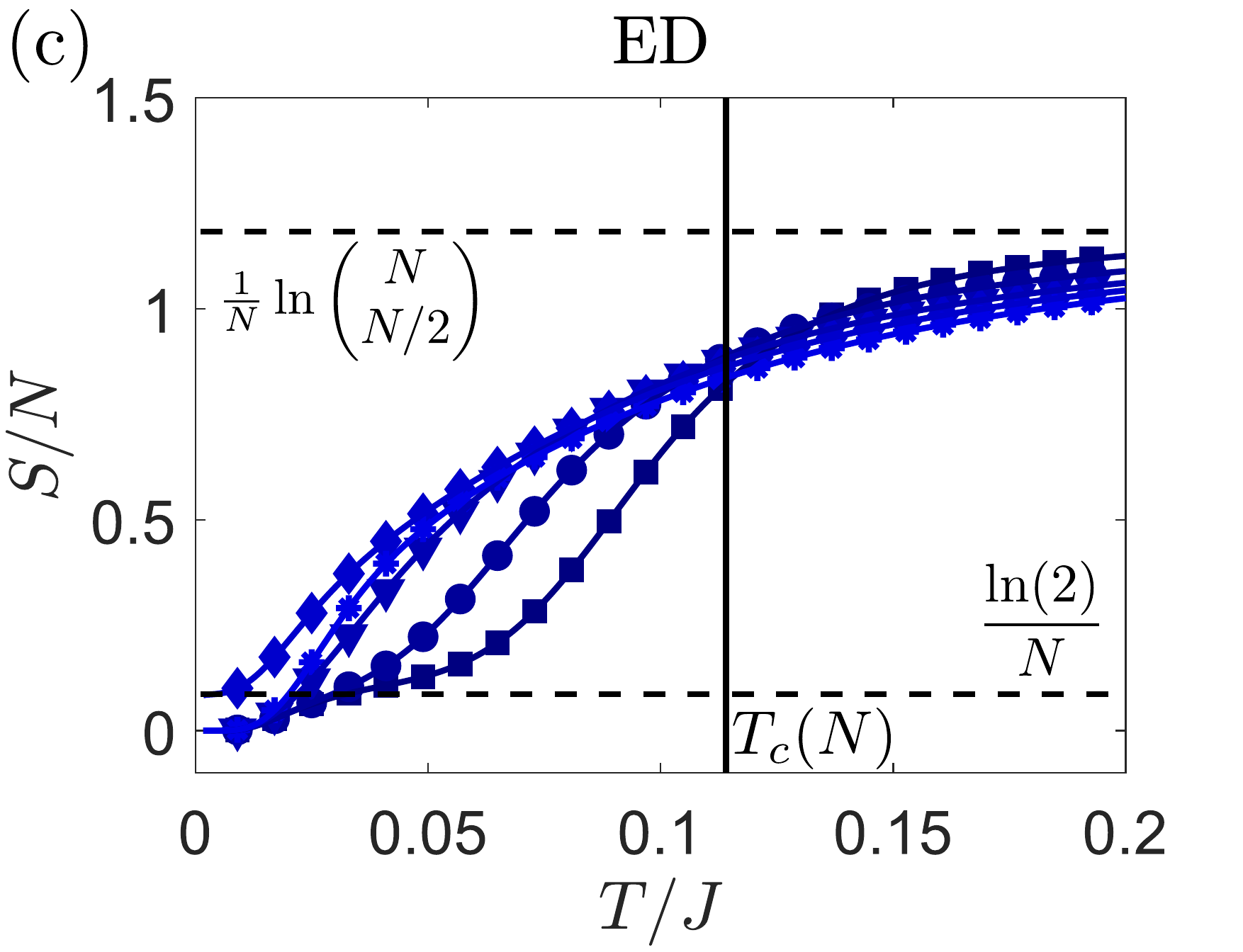}
     \end{minipage}
                 \begin{minipage}{0.32\textwidth}
         \centering
         \includegraphics[width=\textwidth]{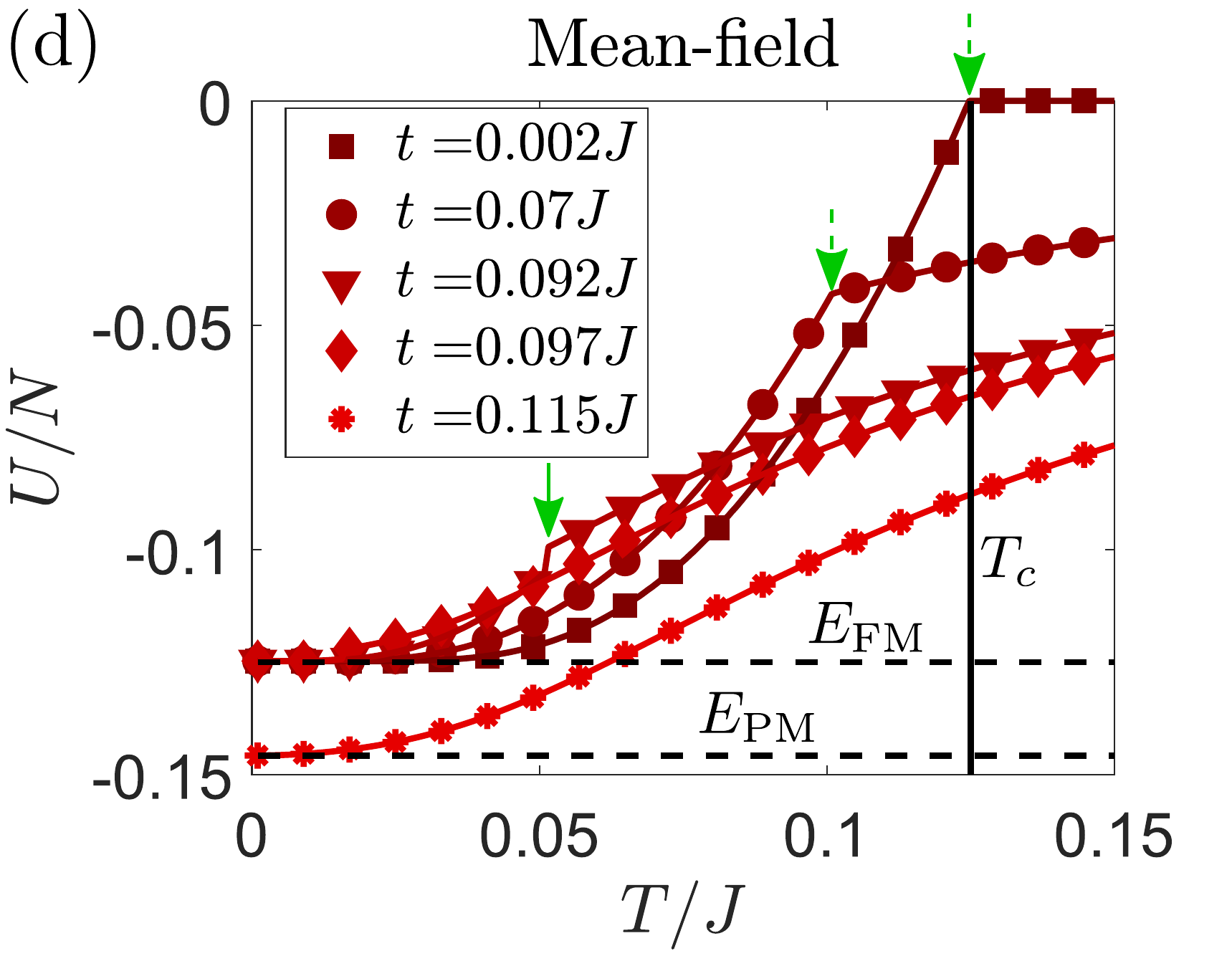}
     \end{minipage}
     \hfill
   \begin{minipage}{0.32\textwidth}
         \centering
         \includegraphics[width=\textwidth]{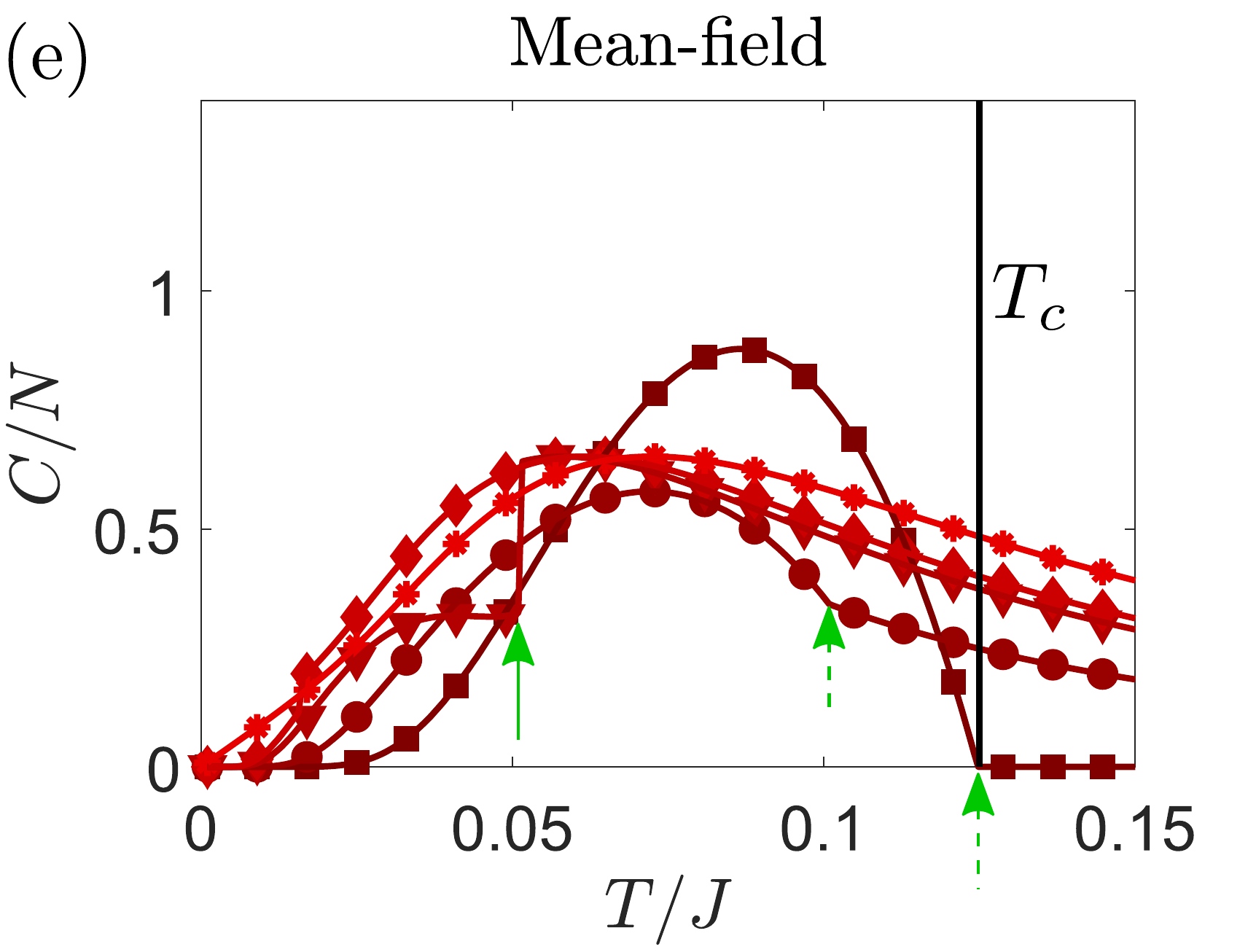}
     \end{minipage}
         \hfill
            \begin{minipage}{0.32\textwidth}
         \centering
         \includegraphics[width=\textwidth]{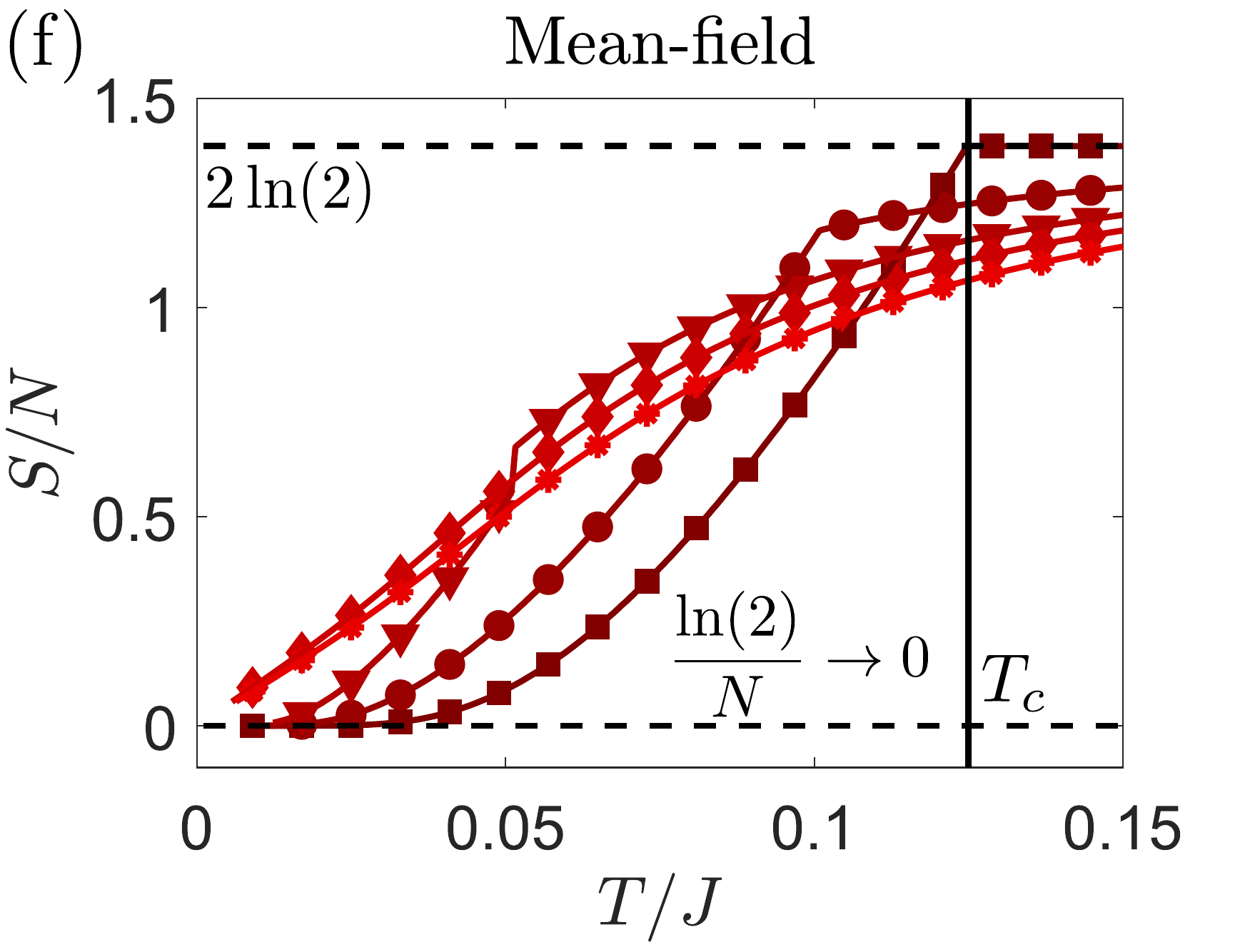}
     \end{minipage}
         \caption{Internal energy density $U/N$, specific heat $C/N$ and entropy density $S/N$ as a function of $T$ at fixed $t$, obtained by ED with $2N=16$ fermions (top row) and by mean-field theory (bottom row). The values $t=0.104J$ (ED) and $t=0.097J$ (MF) correspond to the QCP. Lines for $t=0.092J$ (triangles) show a first order phase transition, highlighted by the green solid arrow. Lines for $t=0.002J,0.07J$ (squared, bullets) show a second order phase transition, highlighted by green dashed arrows. }
      \label{fig: thermodynamics fun of T}
\end{figure*}

We generalize the mean-field approach to finite temperature. In fact, the knowledge of the diagonal Hamiltonian in Eq.~\eqref{eq:diagonal mean field H} allows us to compute the canonical partition function at inverse temperature $\beta$
\begin{align}
	Z = e^{-\frac{\beta}{2} J\mathcal{M}^2(N-1)}\,\prod_{k,\pm}\left(1+e^{-\beta\varepsilon_\pm (k,\mathcal{M})}\right).
	\label{eq:partition_function}
\end{align}
From the partition function the free energy is obtained as
\begin{align}
		F=\frac{J\mathcal{M}^2}{2}(N-1)-\frac{1}{\beta}\sum_{k,\pm}\ln\left(1+e^{-\beta\varepsilon_\pm(k,\mathcal{M})}\right).
	\label{eq:free_energy}
\end{align}
In order to study the finite temperature phase diagram it is useful to consider an expansion of the free energy to fourth order in the order parameter 
\begin{align}
	F = F_0 + F_2 \mathcal{M}^2+F_4\mathcal{M}^4+\mathcal{O}(\mathcal{M}^6),
	\label{eq:F_expansion}
\end{align}
with coefficients
\begin{align}
	&F_0 = -\frac{2}{\beta}\sum_k\ln(1+e^{-\beta\varepsilon_k(t)}),
	\\
	&F_2= \frac{J(N-1)}{2}\left[1-\frac{\beta J}{8}\frac{(N-1)}{N^2}\sum_k\frac{1}{\cosh^2(\beta\varepsilon_k(t)/2)}\right],
	\\
	&F_4\propto\frac{1}{N}\sum_k\frac{\cosh(\beta\varepsilon_k(t))-2}{\cosh^4(\beta\varepsilon_k(t))}.
\end{align} 
Here, we have introduced the shorthand notation $\varepsilon_k(t) = -2t\cos(k)$. Notice that the free energy contains only even powers of $m$ due to time-reversal symmetry. The second order transition is then identified by the conditions $F_2 = 0$, $F_4>0$. In fact, shown in Fig. \ref{fig:Free energy}(c), for $F_4>0$ the free energy has the double well shape typical to of second order phase transitions. This conditions provide an implicit equation for the phase boundary
\begin{align}
	\frac{\beta J}{8}\int_{-\pi}^{\pi}\frac{dk}{2\pi}\frac{1}{\cosh^2(\beta\varepsilon_k(t)/2)} = 1.
\end{align}
In the limit $t\to 0$ the integral becomes one and we find $(T/J)_{c,t=0} = 1/8$, which is in good agreement with the critical temperature as obtained from the exact solution of the model at zero hopping. Then the phase transition becomes first order when $F_4<0$; in fact in this case, in order to preserve the stability, we need to include also the sixth order in the free energy expansion \eqref{eq:F_expansion}. Consequently, as shown in Fig.~\ref{fig:Free energy} (a), the free energy at the critical point has three minima which become degenerate at the critical point signaling a first order phase transition.
\\
Figure \ref{fig:Free energy}(d) shows the points in the $t$-$T$ plane where $F_2 = 0$ (red line) and $F_4 = 0$ (blue line). The intersection between these two lines determines a tricritical point $(T^*, t^*)$ at which the phase transition passes from second to first order.  Within mean-field theory its location is found to be at 
\begin{align}
	(t^*,T^*)\approx (0.089,0.067)J.
\end{align}
This result further corroborates the simple argument and numerical finite size results, see Sec.~\ref{sec: Finite size analysis tricritical point}.  
Minimizing $F$ with respect to the magnetization $\mathcal{M}$, we obtain the same self consistent equation for the order parameter as in the zero temperature case \eqref{eq:selfconsistent_m}, with the ground state expectation values replaced by thermal averages subject to a Fermi-Dirac distributions
\begin{align}
	n_{k,\pm}^{\mathrm{th}} = \frac{1}{1+e^{\beta\varepsilon_{\pm}}(k,\mathcal{M})}.
\end{align}
Fig. \ref{fig: emf} (b) shows the order parameter as a function of $t/J$ and for different values of the temperature around the tricritical point $T=T^*$. For $T<T^*$,  $\mathcal{M}$ displays a discontinuous jump at the transition point, which is then of first order. As the temperature increases the discontinuity becomes smaller reaching zero at the tricritical temperature $T=T^*$. Then for $T<T^*$ the order parameter becomes a continuous function of $t/J$, thus undergoing a second order phase transition. 
\\
Finally the complete mean-field phase diagram, obtained by numerically minimizing the free energy (\ref{eq:free_energy}) with respect to the magnetization, is shown in Fig.~\ref{fig: Phase diagram} (b). Very good agreement is found with the exact numerical phase diagram Fig.~\ref{fig: Phase diagram} (a).  
\section{Thermodynamics}
\label{sec: thermodynamics}
\begin{figure*}[t!]
   \begin{minipage}{0.32\textwidth}
         \centering
         \includegraphics[width=\textwidth]{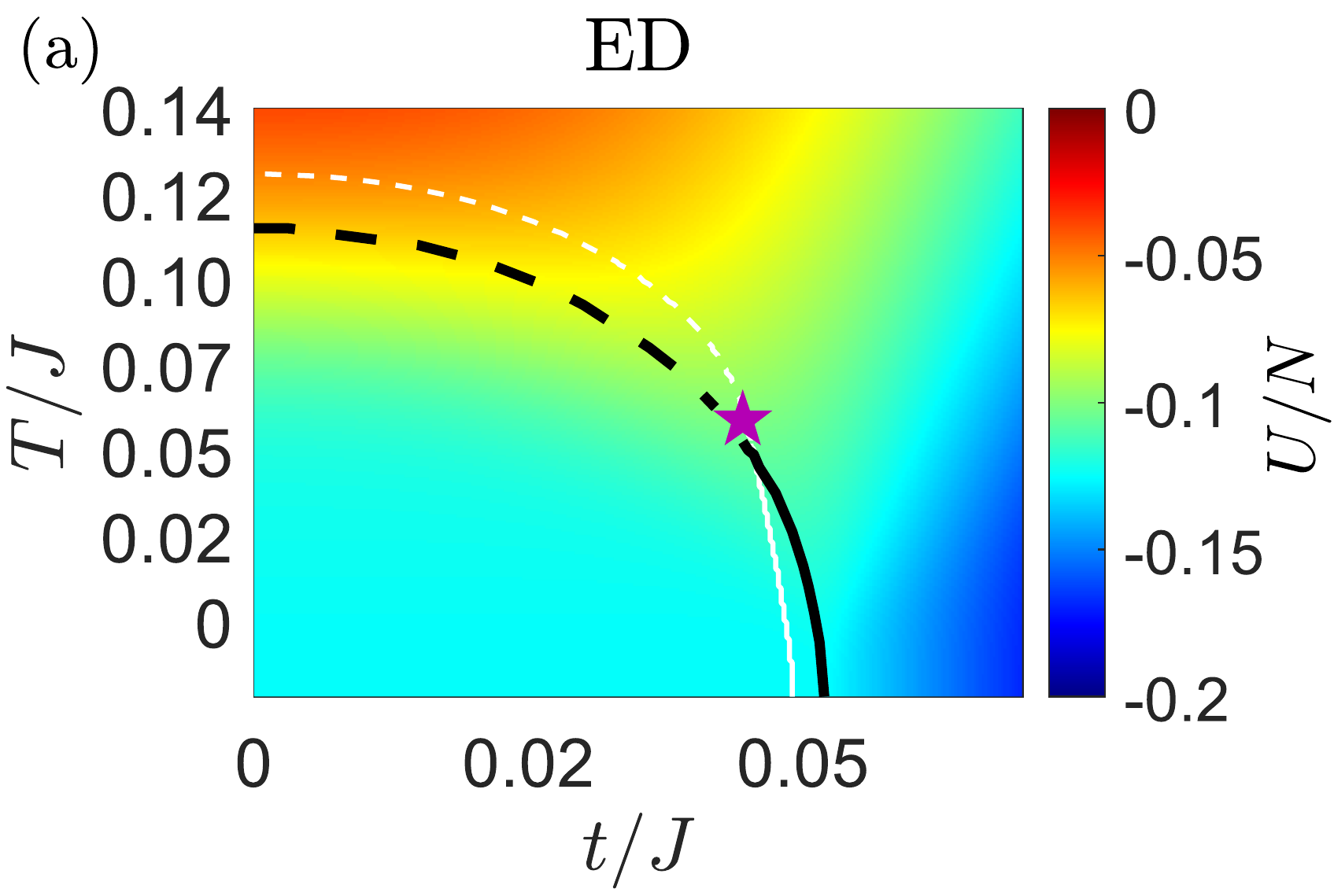}
     \end{minipage}
               \hfill
    \begin{minipage}{0.32\textwidth}
         \centering
         \includegraphics[width=\textwidth]{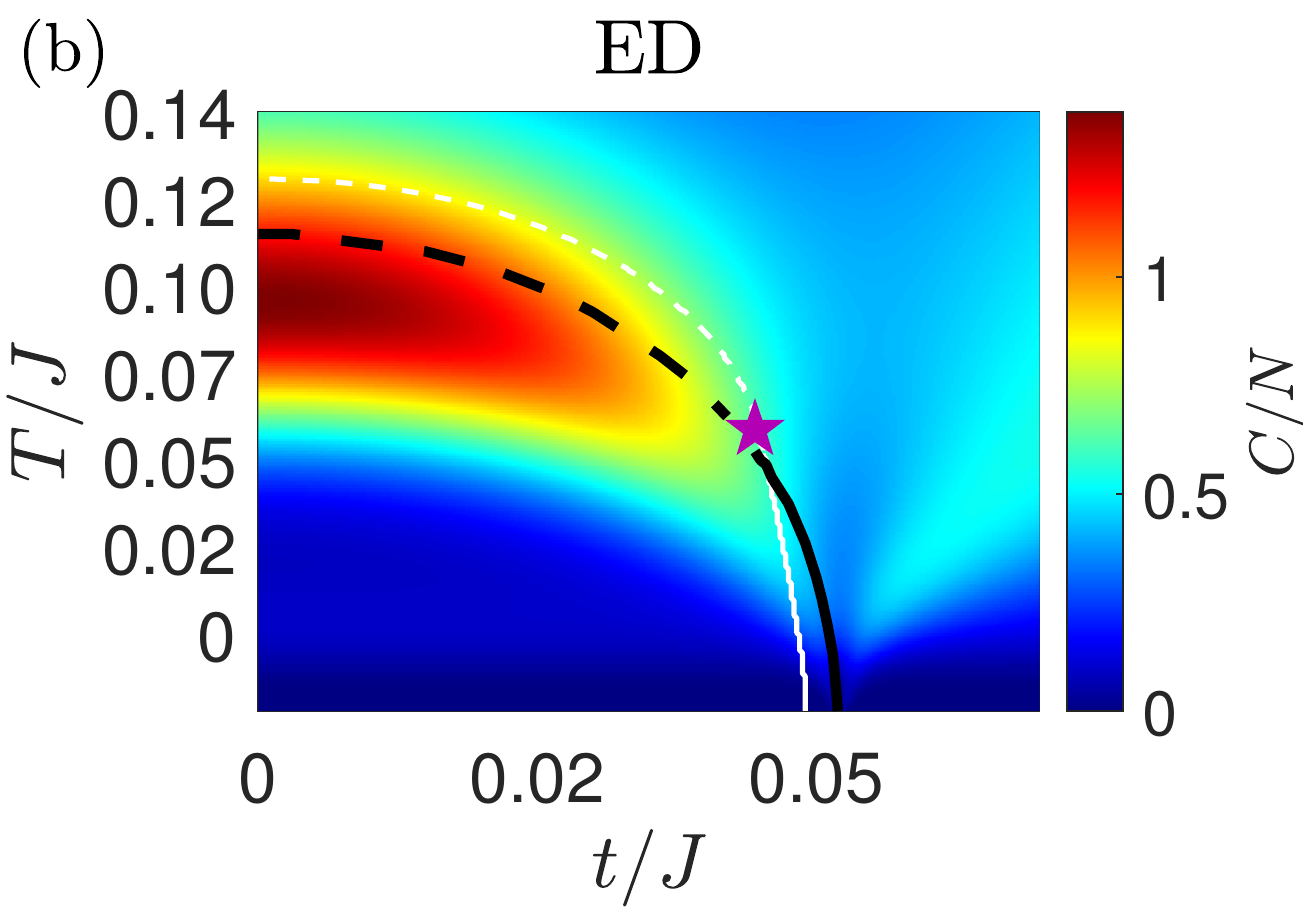}
     \end{minipage}
         \hfill
    \begin{minipage}{0.32\textwidth}
         \centering
         \includegraphics[width=\textwidth]{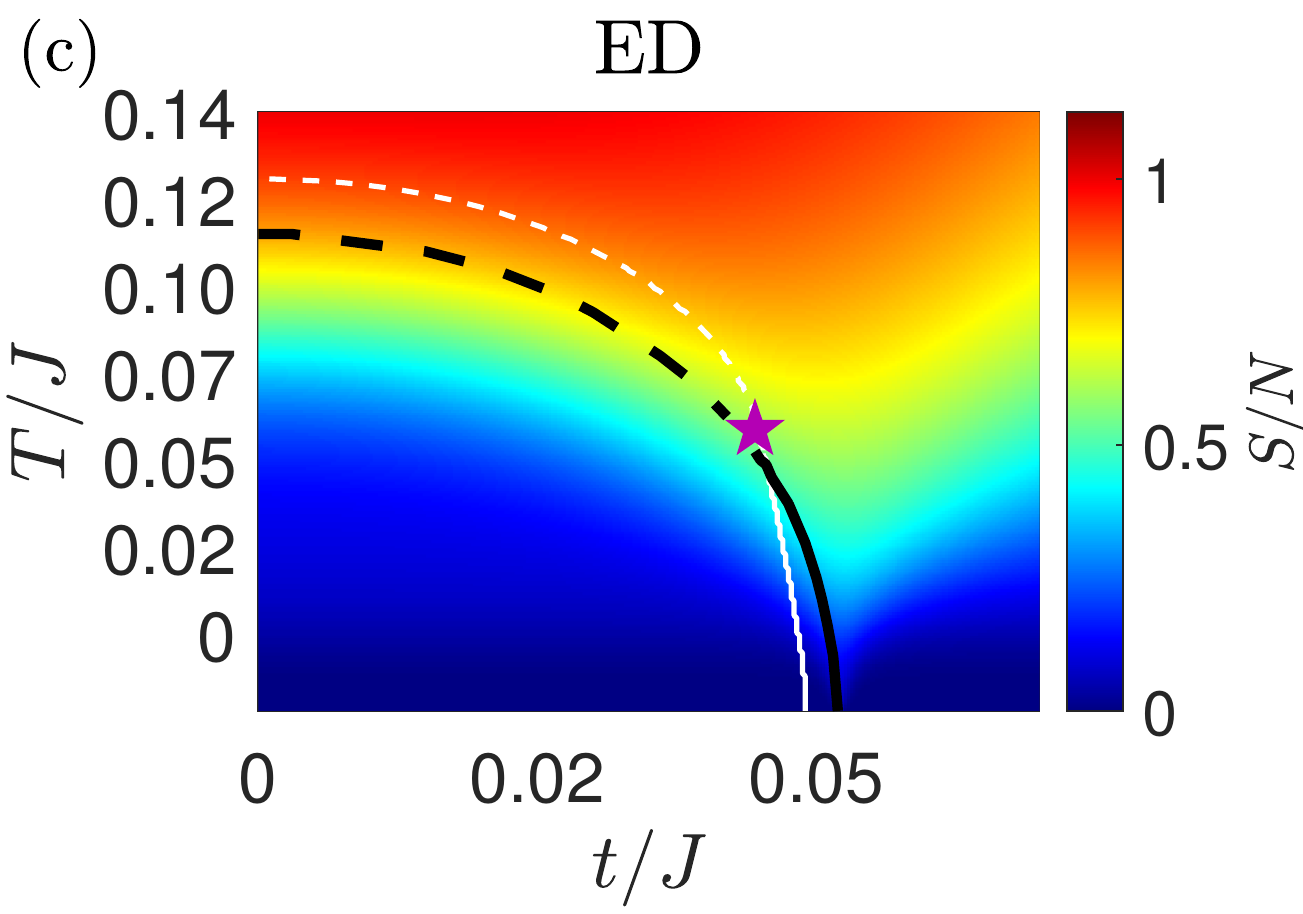}
     \end{minipage}
        \begin{minipage}{0.32\textwidth}
         \centering
         \includegraphics[width=\textwidth]{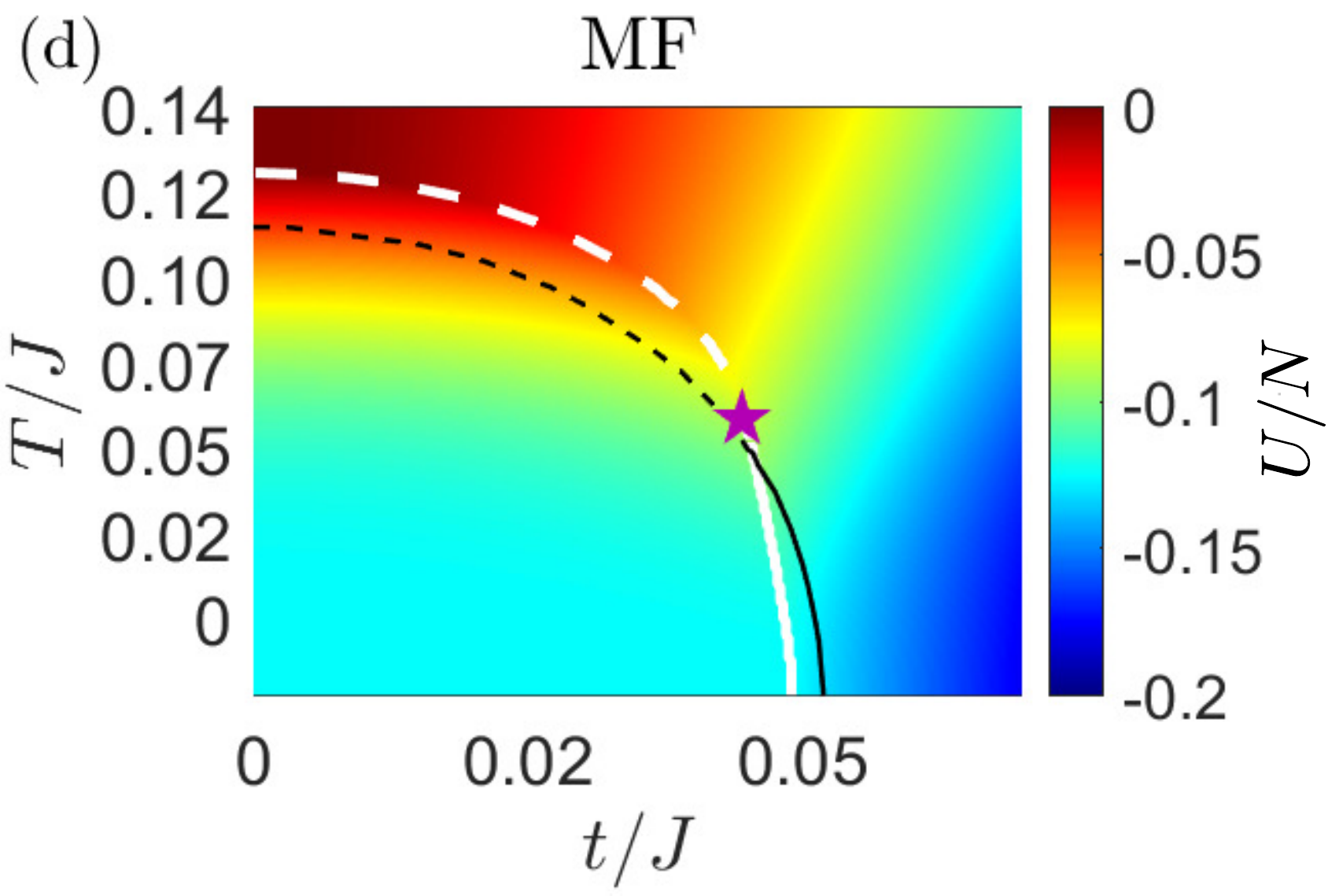}
     \end{minipage}
               \hfill
    \begin{minipage}{0.32\textwidth}
         \centering
         \includegraphics[width=\textwidth]{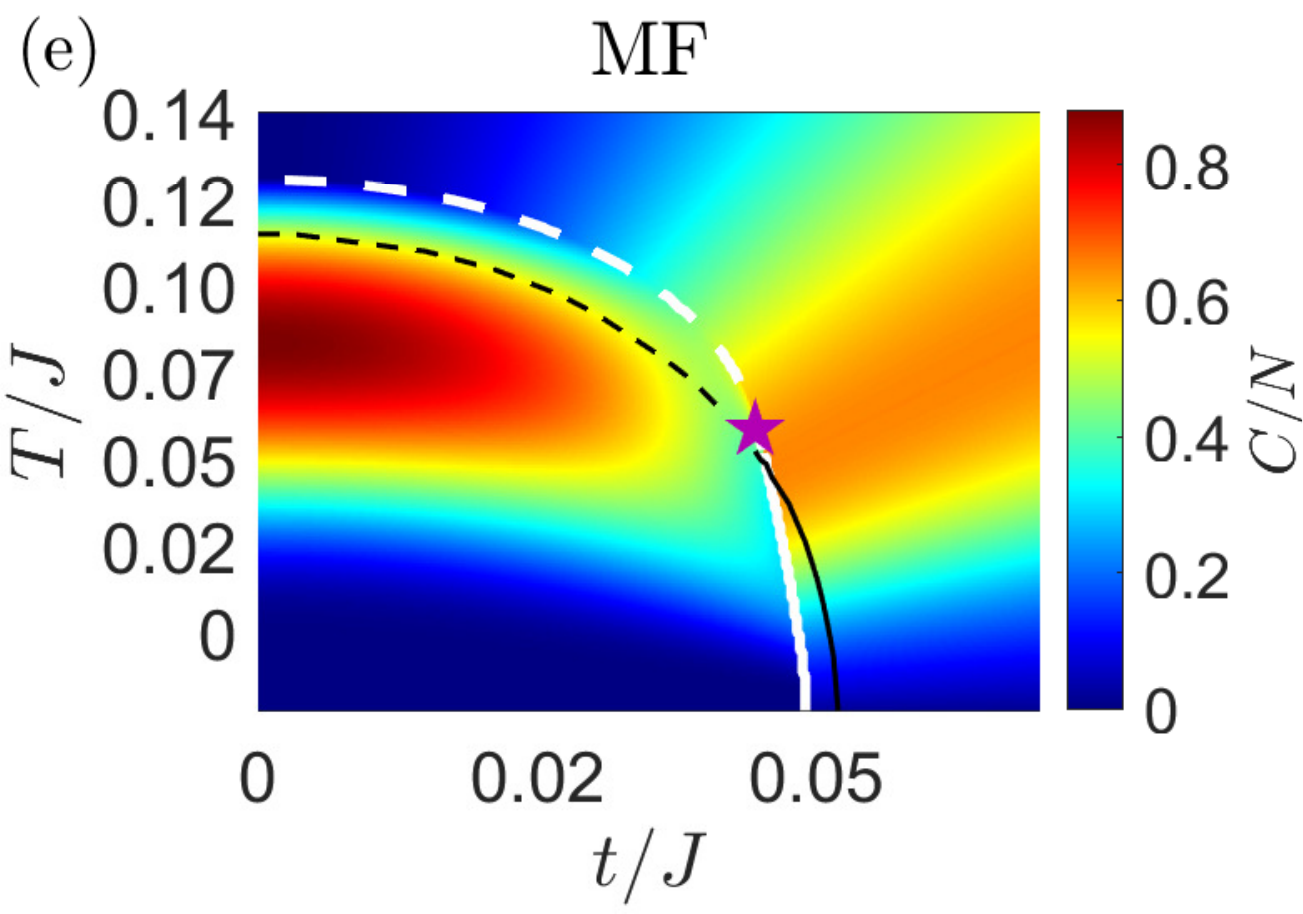}
     \end{minipage}
         \hfill
    \begin{minipage}{0.32\textwidth}
         \centering
         \includegraphics[width=\textwidth]{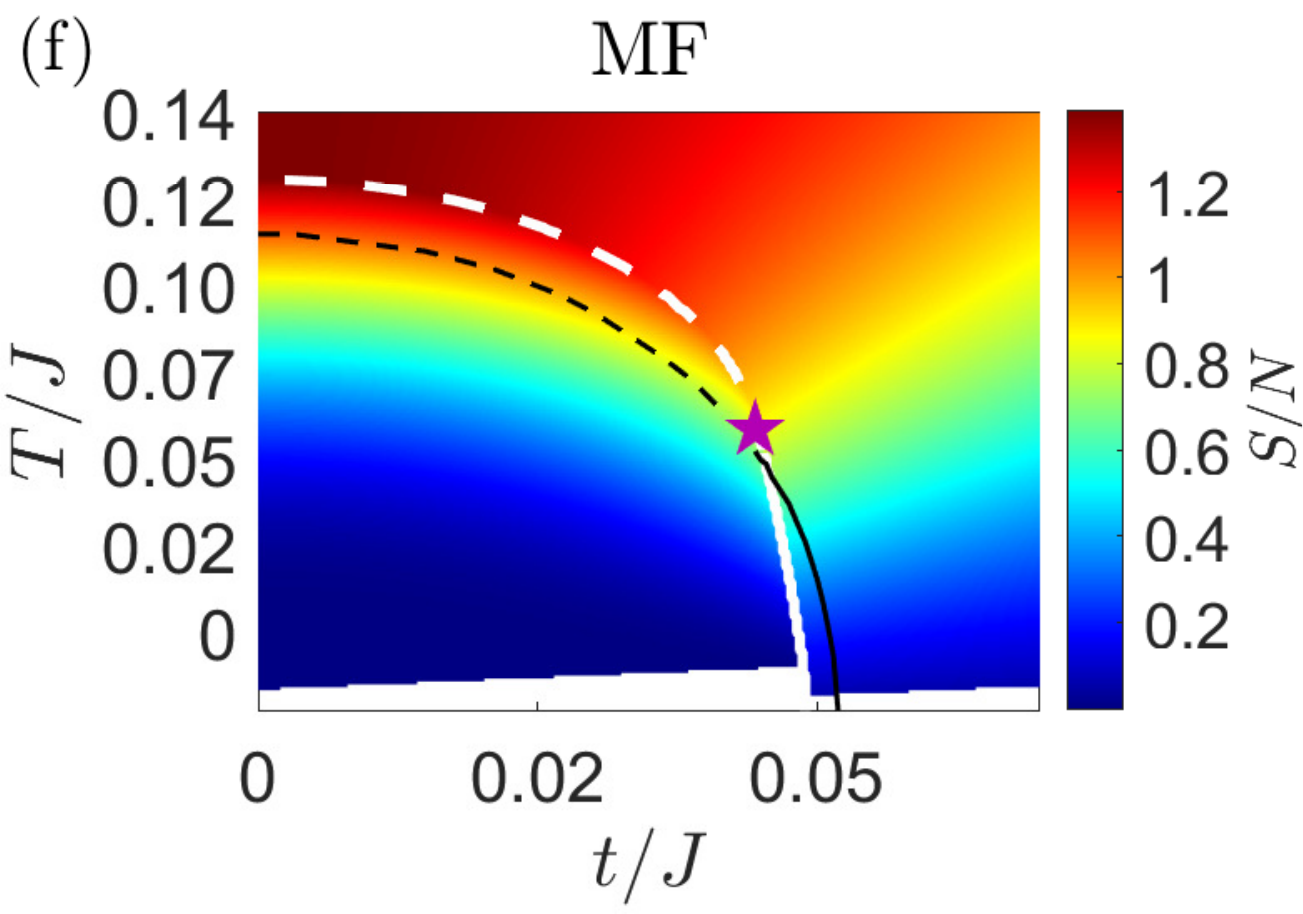}
     \end{minipage}
         \caption{Surface plots for the internal energy density  $U/N$, heat capacity $C/N$ and entropy density $S/N$  for ED (a)-(c) and MF theory (d)-(f). Phase boundaries are obtained from the magnetization and are colored in black (ED) and white (MF). First order phase boundaries (solid lines) and second order phase boundaries (dashed lines) meet at the TCP, as indicated by a purple star.}
    \label{fig: thermodynamic phase diagrams}
\end{figure*}
We investigate the internal energy, specific heat and the entropy
\begin{align}
U=\langle H \rangle, \qquad C=\dfrac{\partial U}{\partial T}, \qquad  S=\ln(Z)+\dfrac{ U}{ T}
\end{align}
of the fermionic quantum HMF in the canonical ensemble. The thermodynamic quantities permit further investigation of the critical points and phase boundaries, as well as a comparison between numerics and MF theory. 
\\
Fig.~\ref{fig: thermodynamic phase diagrams} displays surfaces of the thermodynamic quantities in the $T$-$t$ plane, obtained by ED and MF theory. Both approaches are in seemingly good agreement, especially in the ordered phase and for low temperatures. The QCP is visible in all plots in terms of a cusp ($U$), jump ($C$)  and a non-zero value of $S$. $t_c$ is somewhat larger in the numerics than in MF theory. The CCP is most pronounced in the vicinity of the maximum in $C$; here, the numerics has a lower value of $T_c$ than in MF theory. Signatures of the TCP are already visible in MF theory for the heat capacity (Fig.~\ref{fig: thermodynamic phase diagrams} (e)) indicated by the maximum that builds of from the top right corner of the PM phase. For the numerics, the TCP remains hidden in Fig.~\ref{fig: thermodynamic phase diagrams}; later we will extract more information from the surface plots by taking generalized derivatives in parameter space.  
\\
We proceed to a more detailed and quantitative treatment in Fig.~\ref{fig: thermodynamics fun of T} which shows $U/N$, $S/N$ and $C/N$ for fixed, representative values of the hopping as a function of $T$. The temperature dependence of the thermodynamic quantities differs strongly depending on whether the system has few charge fluctuations ($t\approx 0$), moderate charge fluctuations ($t<t^*<t_c$) or large charge fluctuations ($t_c<t$). 
\subsubsection{Few charge fluctuations $t\approx 0$}
\begin{figure}[t!]
         \centering
         \includegraphics[width=0.5\textwidth]{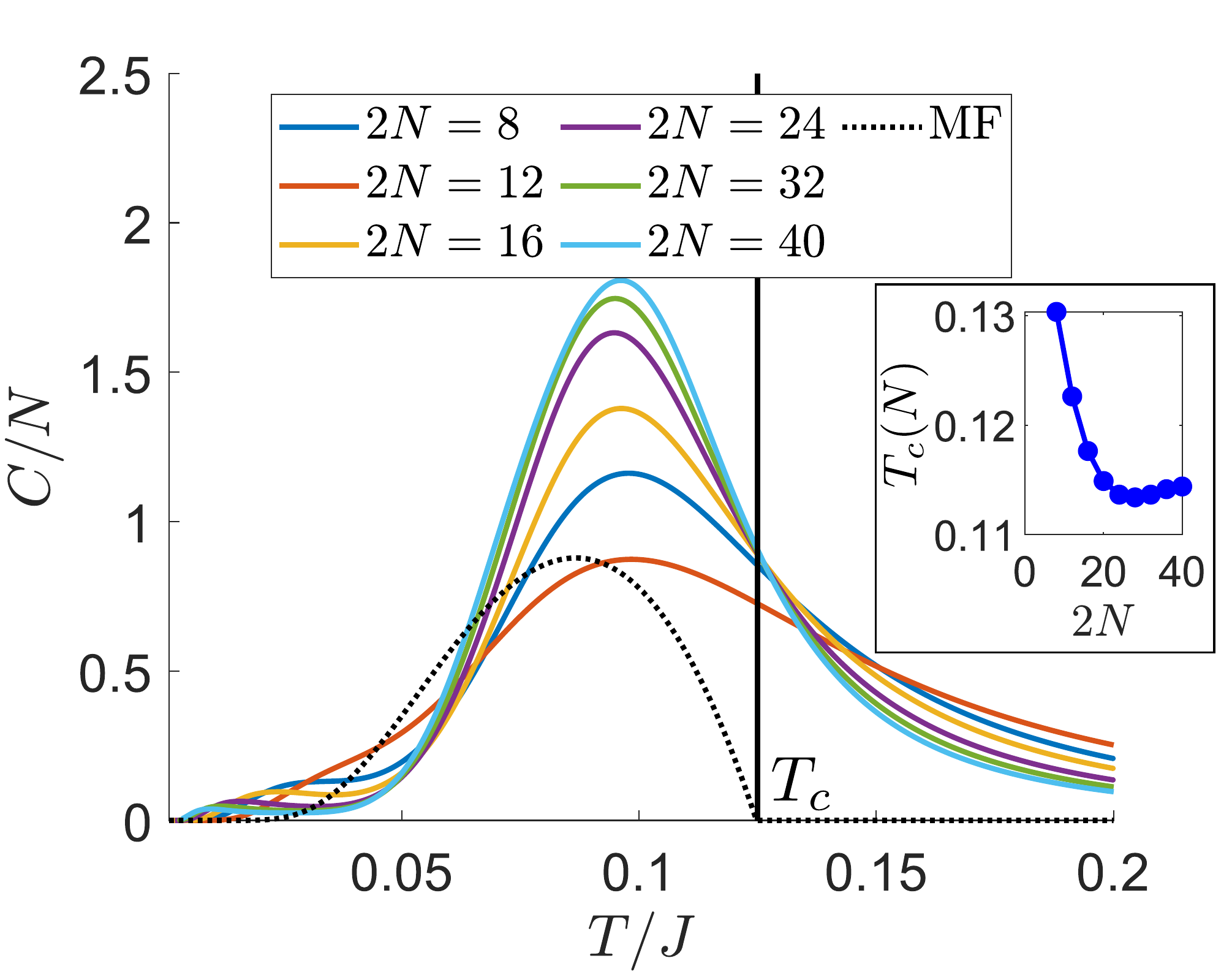}
         \caption{Heat capacity for different system sizes at zero hopping, obtained by counting degeneracies as explained in App.~\ref{app: zero hopping}. The critical temperature $T_c$ is obtained by calculating $\mathrm{min}\{\partial_T C\}$ and is in agreement with the the mean-field value $T_c/J=1/8$.}
      \label{fig: heat capacity zero hop}
\end{figure}

In the ordered phase and with few charge fluctuations ($t\approx 0$), $U$ remains nearly constant in the low-temperature regime $T \lessapprox 0.05J$; also $C$ is very flat and $S$ grows slowly. This reflects that the manifold of high-energy states is gapped out due to the dominance of interactions over hopping in the full quantum HMF; in MF theory the lower band contributes with constantly low density of states at half-filling to $U$ and agrees well with the numerical result. For $T >0.05J$, $U$ increases then significantly, indicated by the maximum of the heat capacity: this occurs due the onset of the highly degenerate high-energy states in the many-body spectrum  for $T\sim\mathcal{M}$, see Fig.~\ref{fig: spectrum} (a). 
\\
In MF theory this is due to increase of the density of states away from the middle of the band. The second order phase transition is, however, only clearly visible in MF theory and appears as a cusp in $C$, as indicated by a dashed green arrow. In the numerics, the CCP is observed after performing a finite size scaling, see Fig.~\ref{fig: heat capacity zero hop}, and approaches the MF value. Notice that the entropy is a convex function in the ordered phase for $0.05J<T<T_c$ and concave for $T>T_c$, and approaches the limit $S\to 2N\ln(2)$ at high temperatures.
\\

\subsubsection{Moderate charge fluctuations $t^*<t<t_c$}
%

%
For moderate charge fluctuations we observe significant ramifactions of the described behavior. This happens due to the hopping-induced splitting of the high-energy states, which supports the formation of a quasi-continuum in the many-body spectrum, separated to the gapped ground state, see Fig.~\ref{fig: spectrum}. After $T$ surpasses the hopping-reduced gap, $U$ rises approximately linear in temperature. $S$ increases drastically and acquiring a concave shape for much lower temperatures. For hopping strengths $t^*<t<t_c$, $U$ exihibits then a discontinuous jump in MF theory, marked by a solid green arrow, which signals a first order transition; this is in stark contrast to the discontinuity in $C$ for $t<t^*$ signaling a second order transition; there is a latent heat in the system to be compensated before going through the phase transition for $t^*<t<t_c$. Due to the finite-size effects this is not clearly visible in the ED data of Fig.~\ref{fig: thermodynamics fun of T} without further analysis, see Sec.~\ref{sec: Finite size analysis tricritical point}.
\\
The QCP ($t_c=0.104J$ for $2N=16$ fermions) appears directly in the internal energy, which is reduced in the disordered phase at zero temperature, since $E_\mathrm{PM}<E_\mathrm{FM}$. In the vincinity of the QCP, i.e. $(t,T)=(t_c\pm \epsilon,\epsilon)$, the system may be regarded as an effective two-level system with partition function $Z(\beta)\approx e^{-\beta E_\mathrm{FM}}[1+e^{\beta( E_\mathrm{FM}- E_\mathrm{PM})}]$. Hence, the heat capacity displays an enhanced value at $t\approx t_c$ as observed in the data, see also inset of Fig.~\ref{fig: thermodynamics fun of T}.  The entropy has a  non-zero value $S=\ln(2)$ at the QCP a clear signature for a two-level system. Both features are absent in the plots of the MF data due to the thermodynamic limit. 
\\

%
\begin{figure*}[t!]
	\begin{minipage}{0.32\textwidth}
		\centering
		\includegraphics[width=\textwidth]{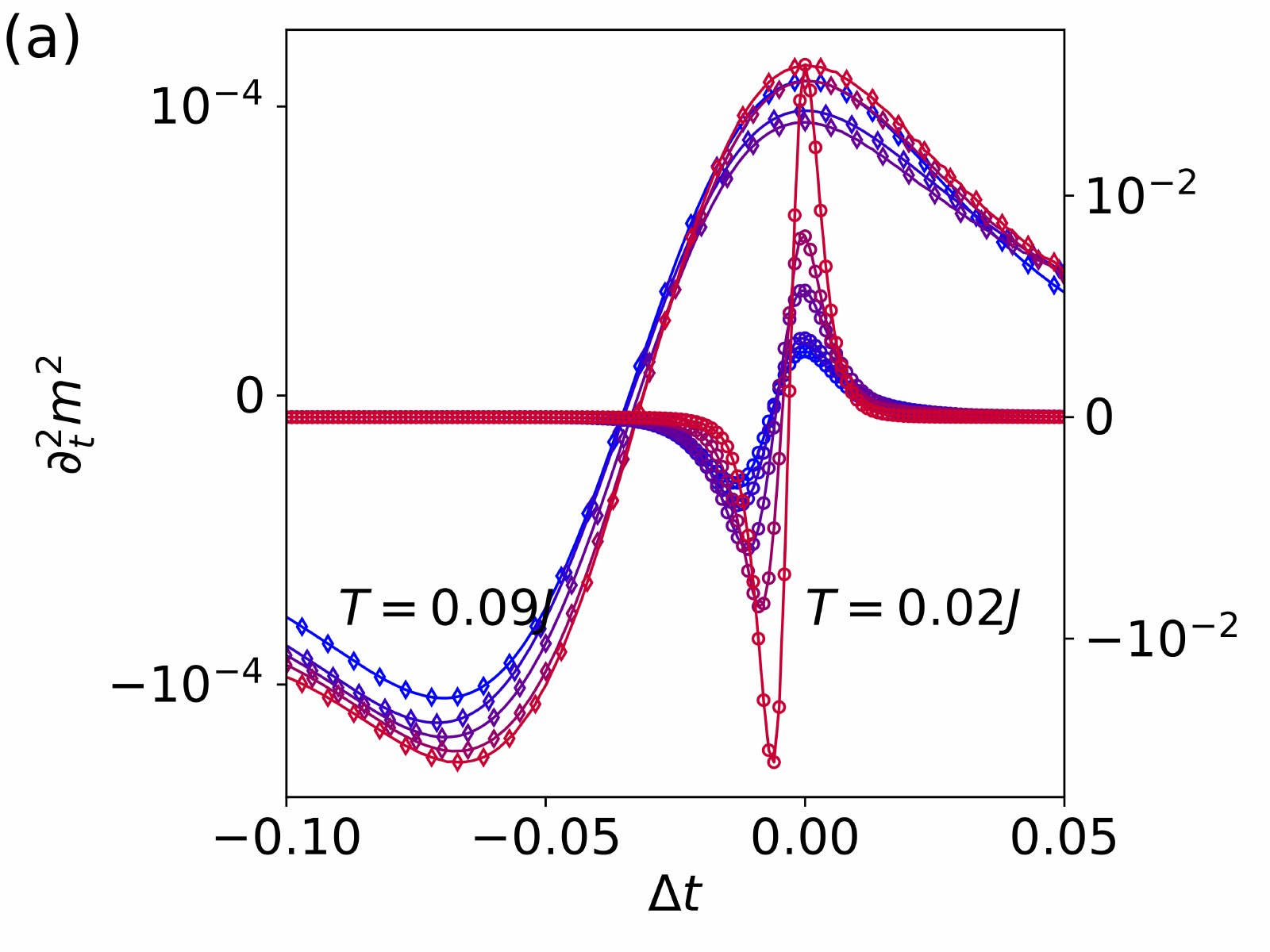}
	\end{minipage}
	\begin{minipage}{0.32\textwidth}
		\centering
		\includegraphics[width=\textwidth]{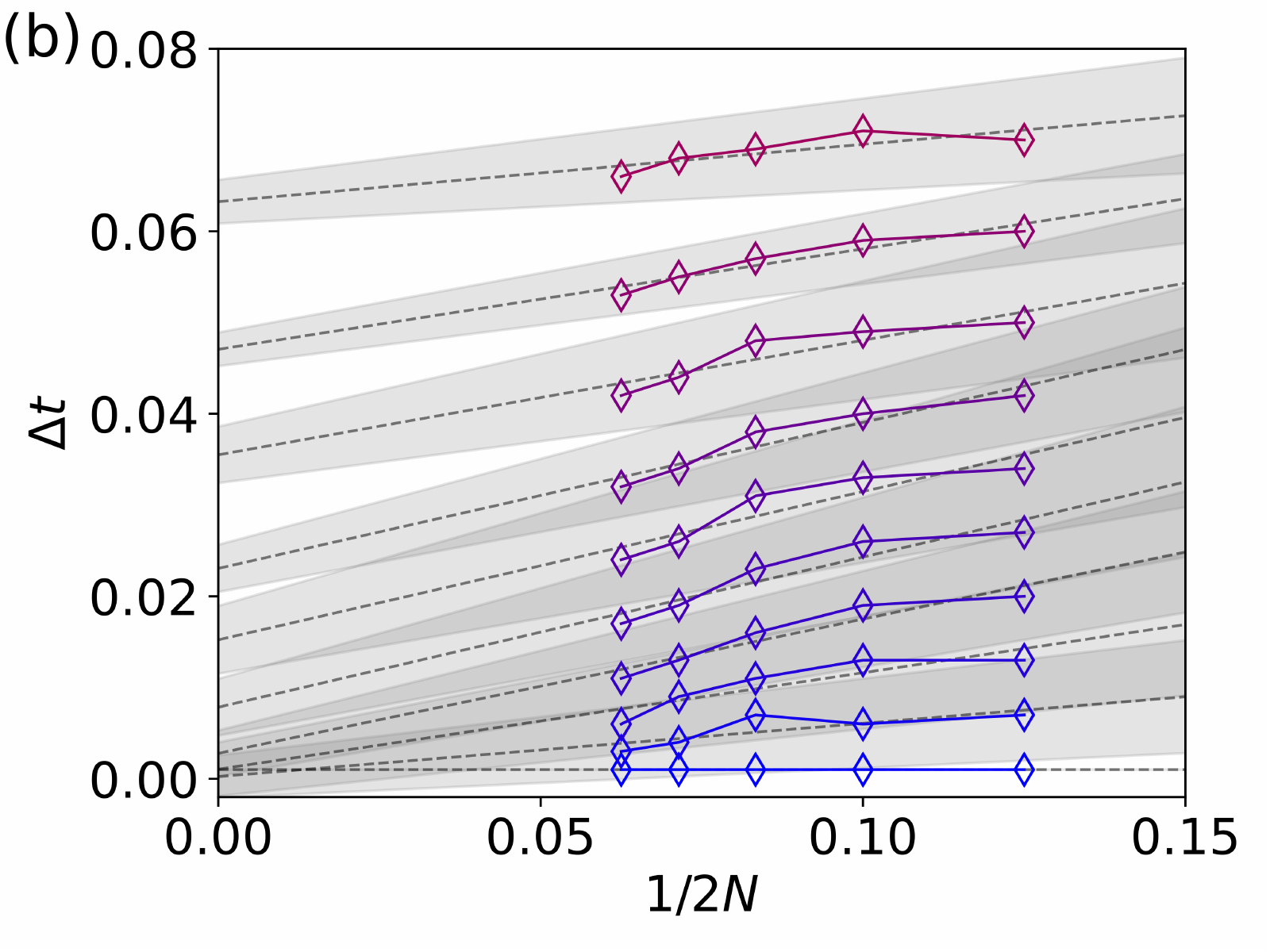}
	\end{minipage}
	\begin{minipage}{0.32\textwidth}
		\centering
		\includegraphics[width=\textwidth]{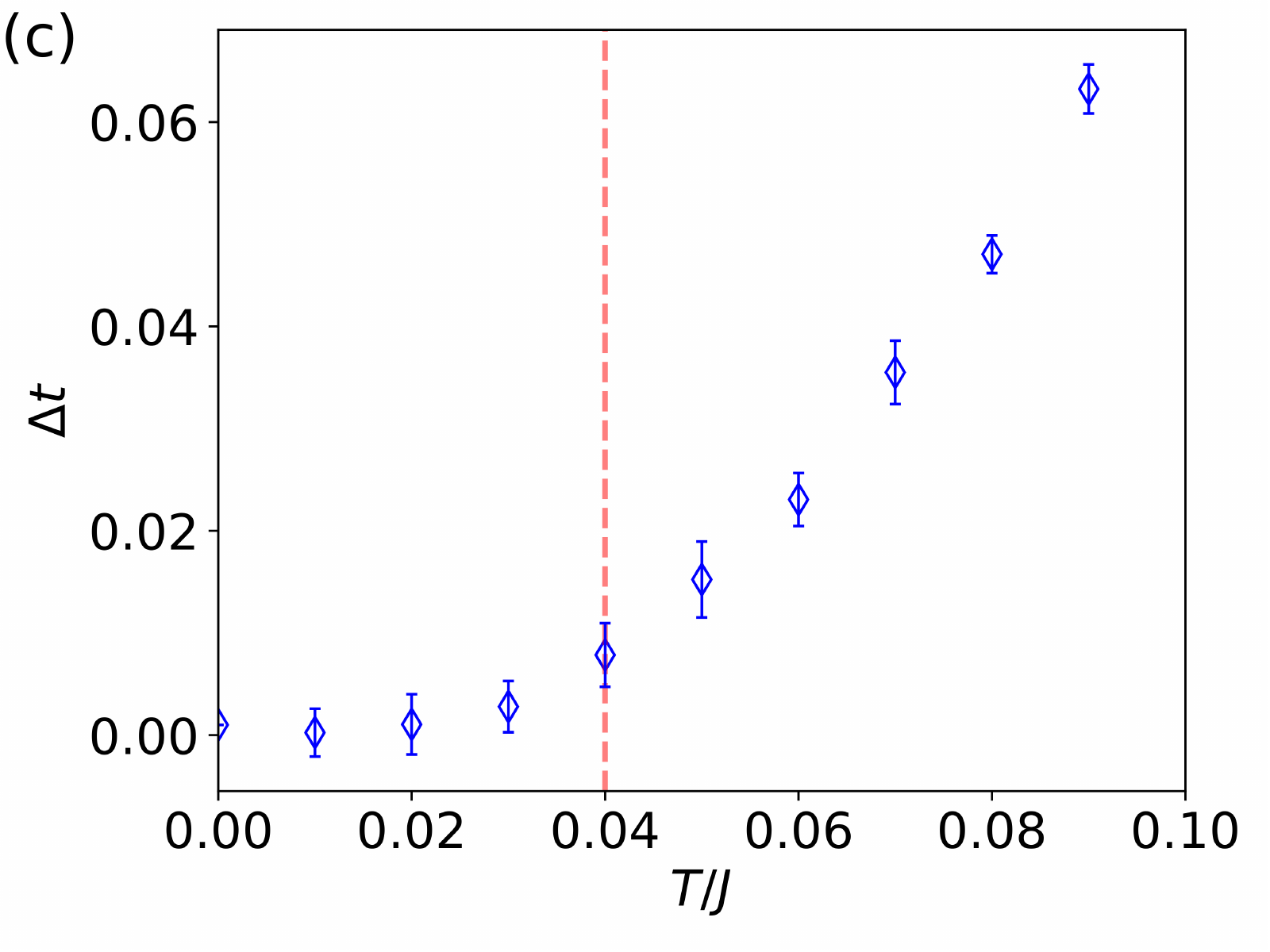}
	\end{minipage}
	\caption{Finite size analysis of the transition window for the ED data. (a) Cut along the axis of hopping of the second partial derivative of the magnetization for different temperature values. System sizes increase from $2N=8$ (red) to $2N=16$ (blue). The left vertical axis corresponds to data at $T=0.09J$, the right to $T=0.2J$. The origin of both data sets is set to the point of maximum curvature $t_a$. (b) Size of transition window $\Delta t=(t_a-t_b)/J$ (as defined in Sec.~\ref{sec: Finite size analysis tricritical point}) for different system sizes $N$ and temperatures $T=0$ (bottom) to $T=0.9$ (top) in steps of $\Delta T=0.1$. The dashed black lines are linear fits extrapolating the finite size data to smaller $1/N$. The shaded areas correspond to the square roots of the covariance of the fits and give an estimate of the errors of the fitting procedure. (c) $1/N\to 0$ extrapolation for the temperatures shown in (b).}
	\label{fig: finite size tricritical}
\end{figure*}

\section{Finite size analysis of the tricritical point}
\label{sec: Finite size analysis tricritical point}
We analyze the ED data in more detail,
and provide solid evidence for the existence of the tricritical point and a first order line in the phase diagram.  It is not easy to visually identify a tricritical point in the magnetization for the system sizes available (up to $2N=16$ fermions), compare Fig.~\ref{fig: Phase diagram}. Therefore, a more involved finite size analysis is needed. In particular, we will look at different cuts along the $t$-axis in the phase diagram Fig.~\ref{fig: Phase diagram} (a). 
\\
In the thermodynamic limit, a first order transition is identified  by a discontinuous jump in $m^2$, while a second order transition is identified  by a cusp in $m^2$. For any finite $N$ and at non-zero temperature, $m^2$ is a continuous function, and the transition happens within a finite window of hopping values. In the following we make use of the different behavior of the window size in the thermodynamic limit to determine the order of the phase transition by extrapolation from our finite size data. Concretely, the window size goes to zero as $N\to \infty$ for a first order transition, while it remains finite for a second order transition. We define the size of this window $\Delta t =(t_a-t_b)/J$ by taking the difference at designated hopping values $t_b$ and $t_a$  before and after the transition at which the curvature of the order parameter $m^2$ becomes extremal, i.e. we determine $\partial_t^2 m^2(t)\mid_{t_{a,b}}=0$. 
\\
Fig. \ref{fig: finite size tricritical} (a) shows $\partial_t m^2$ as a function of the window $\Delta t$ for qualitatively different temperatures $T<T^*$ and $T>T^*$. It is clearly visible that this quantity decreases drastically with increasing system sizes in the first case ($T<T^*$), while it stays nearly constant in the second case ($T>T^*$). This indicates  a qualitative difference between the two scenarios where the former corresponds to a first order and the latter to a second order phase transition. 

In Fig. \ref{fig: finite size tricritical} (b) a finite size extrapolation of the transition window size is shown. Here, we fit the finite size values in $1/N$ linearly and extract the projected window size as $1/N\to0$. Due to errors from the small available system sizes, we include the covariance of the fitting procedure as gray shaded areas. Despite of uncertainties, it is clearly visible that for $1/N\to 0$ the transition window goes to zero within the errorbars for low temperatures, while it approaches a finite value for large temperature. This trend occurs still within the phase boundaries of the ordered phase.

Figure \ref{fig: finite size tricritical} (c) shows the $1/N\to 0$ extrapolation for the size of the transition window with respect to temperature.
Here it becomes apparent that for small temperatures $T<0.04J$ the extrapolation is essentially zero, indicating a first order phase transition; for larger temperatures a transition to a second order one is observed. The temperature value at which this qualitative change is observed can be identified with the tricritical point at which the mean field calculation shows a transition from a first to second order line.  

The extrapolation procedure suggests a numerical value close to $T^*_\text{ED}=0.04J$. The discrepancy with the mean field result $T^*_\text{MF}=0.067J$  can be explained by the lack of large system sizes for the fitting procedure, as well as effects due to quantum fluctuations not included in the mean-field treatment. 
\section{Conclusion}
\label{sec: conclusion}
In this paper, we have proposed a quantum mechanical generalization of the fully connected Hamiltonian mean-field model \cite{Antoni1995}. The classical motion of $N$ fully coupled rotators is realized in the quantum case by an all-to-all $X$-$Y$ interaction among spin-$\frac{1}{2}$ fermions, that hop on a one-dimensional ring. For definiteness, we have restricted our analysis to ferromagnetic coupling and half-filling. In this scenario, the hopping on the lattice can be viewed as charge fluctuations of the ordered ground state, which compete with the ferromagnetic $X$-$Y$-type order, and drive the system into the disordered phase. The main finding of the paper is the tricritical point at non-zero temperature and hopping, separating a first from a second order phase transition.
\\
The phase diagram of the model is studied as a function of temperature and hopping with exact diagonalization complemented by mean-field theory. We find a first order quantum critical point at zero temperature and non-zero critical hopping, and a second order  classical critical point at zero hopping and non-zero critical temperature. At the quantum critical point the ground state changes abruptly from a state with maximum total spin (ferromagnet) to a state with minimum total spin (paramagnet). We have shown how the level crossing can be understood from basic symmetry arguments. MF theory and numerics agree on the determination of the quantum critical point.
\\
The phase boundary, which extends to finite temperature from the quantum critical point, remains first order and becomes second order only at a tricritical point. In the MF analysis, the effective model involves particles with spin uniformly aligned in the direction of the magnetization, resulting in two lower-energy bands separated by an order-induced gap. The energetically lower (upper) band is completely filled (empty) at zero hopping; varying the hopping gradually tunes the filling of the bands. At the gap closing the residual occupation difference, defining the magnetization, can take either a finite or a zero value. This determines consequently the order of the phase transition.
\\
The resulting tricritical point $T^*$ is clearly seen in mean-field theory, with analytic expressions of the free energy $F$ as a function of magnetization \eqref{eq:F_expansion} at hand, with $F$ following textbook Landau theory. In the vicinity of the tricritical point, $F$ is a fourth order polynomial in the magnetization with double-well shape for $T>T^*$, as expected from a second order transition, and a sixth-order polynomial with three minima for $T<T^*$, as expected from a first order transition. We also observe clear signatures of the change in the order of the phase transition in the numerics by finite-size analysis.  A detailed study of thermodynamic observables has been further performed and displays salient features, such as jumps and cusps at the critical points. 
\\
This work sets the ground for future investigations of tricriticality in the quantum regime in the presence of long-range interactions. Further, the study of the dynamics of the quantum HMF model represents a testbed of critical phenomena for experiments on an envisaged long-range quantum computer.
In particular, it would be of great interest to inspect the dynamics of the entanglement
entropy and its potential anomalous resilience far from equilibrium.
\begin{acknowledgments}
We gratefully acknowledge discussions with Giovanni Modugno. HS acknowledges
funding through CRC 183 from the German Research Foundation. JD acknowledges support from the German Research Foundation through project EV 30/12-1 and SFB 1277 and the German Academic Scholarship Foundation. SS wishes to acknowledge funding from the European Research
Council under the Horizon 2020 Programme Grant Agreement n. 739964 ("COPMAT").
This work is part of the MIUR-PRIN2017 project Coarse-grained description for non-equilibrium systems and transport phenomena (CO-NEST) No. 201798CZL. 
\end{acknowledgments}
\newpage
\appendix
%
%

\section{Spectrum for zero hopping case and }
We review some spectral results at zero hopping \cite{BotetJullianPRB1983}. Taking states with homogeneous particles density into account (i.e. one spin per site), the quantum HMF reduces to the fully connected XY-model, also known as the Lipkin-Meshkov-Glick (LMG) model at zero field \cite{lipkin1965}. Using the spin projection $S_{i,a}=\sigma^a_{i}/2$ ($\hbar=1$), we can write the interaction as $H_{J} =-(J/2N)[S_x^2+S_y^2-K_x-K_y]$,
where $S_a=\sum_i S_{i,a}$ is the total spin projection in direction $a$  and $K_a=\sum_i S_{i,a}^2$ the squared euclidean norm. For spin-$\frac{1}{2}$ particles, the norm gives a constant shift $K_{a}=\frac{N}{4}$ since for any Pauli-matrix $(\sigma^a_{i})^2=1$. We can proceed by expressing the Hamiltonian in terms of the total spin $S$ and its projection $S_z$
\begin{align}
\label{eq: Hamiltonian fully interacting}
    \tilde{H}_{J} =- \frac{J}{2N}\left(S^2-S_z^2\right)+\frac{J}{4}.
\end{align}
The spectrum reads
\begin{align}
\label{eq: spectrum fully-interacting}
    \tilde{E}_{J} =- \frac{J}{2N}\left(S(S+1)-S_z^2\right)+\frac{J}{4}.
\end{align}
The magnetization is equal to the interaction energy up to a scaling and a constant via
\begin{align}
 \tilde{E}_J=\frac{J}{4}\left(1-\frac{Nm^2}{2}\right)  
\end{align}

In the ferromagnetic (FM) case $J<0$, the ground state is obtained by maximizing the total spin $S=N/2$ while simultaneously minimizing the spin projection, placing an equal number of spin up and down. In a Fock basis the ground state reads ($\mathcal{N}$ is a normalization)
\begin{align}
    |\mathrm{FM}\rangle = \frac{1}{\sqrt{\mathcal{N}}}\left(|\uparrow\downarrow\uparrow\dots >+\mathrm{"all\,transpositions"}\right),
\end{align}
with eigenvalue $E_\mathrm{FM}=-JN/8$ for even $N$ . 
At zero temperature, this implies a unit magnetization for $J>0$.
\\
In the antiferromagnetic (AFM) case $J>0$, the ground state is obtained by minimizing the total spin while maximizing the total projection. Since $S_z=-S,-S+1,\dots,S$ this is achieved for total spin zero and thus $S_z=0$ ($N$ even) and total spin $S=1/2$ and $S_z=\pm 1/2$ for ($N$ odd). The magnetization vanishes for antiferromagnetic coupling for $N\rightarrow \infty$. 
\\
Expression \eqref{eq: Hamiltonian fully interacting} can be generalized to fermions, when we take into account the possibility of doubly occupied sites. Due to the conservation of local variance in the particle number at $t=0$, the 
Hamiltonian separates into blocks with equal number of doubly occupied sites. Let us denote the number of doubly occupied sites by $N_{\#}$. The spectrum at $t=0$ is given then given by
\begin{align}
\label{eq: fully interacting spectrum} 
    E_J=- \frac{J}{2N}\left(S(S+1)-S_z^2\right)+\frac{J}{4}\frac{N-N_\#}{N}.
\end{align}

\section{Thermodynamics for zero hopping}
\label{app: zero hopping}
We give some details of the calculation of the thermodynamic quantities at zero hopping in the canonical ensemble. The main difficulty is to determine the degeneracies $g$ which enter the partition function (with $\beta=(k_B T)^{-1}$)
\begin{align}
Z = \sum_{N_{\#}}\sum_{s} g(N_{\#},s)\sum_{s_z}e^{-\beta E(N_{\#},s,s_z)}.
\end{align}  
The partition function is summed over the spin-projections $S_z=-S,...,+S$, total spins $S$ and the sectors with different number of doubly occupied sites. There are three sources of degeneracies in the spectrum:

i) Eigenvalues with different $s,s_z,N_\# $ can be equal.

ii) A chain of spin-$\frac{1}{2}$ fermions combines to degenerate total spins $s$. Iteration of the   the rules of angular momentum addition leads to direct product decomposition \cite{Cirac1999}
\begin{align}
\bigotimes_{k=1}^n\mathbf{\frac{1}{2}}= \bigoplus_{k=0}^{\lfloor n/2 \rfloor}\bigg(\frac{n+1-2k}{n+1}
\begin{pmatrix}
n+1\\
k
\end{pmatrix}
\bigg)(\mathbf{n+1-2k}).
\end{align}

iii) Double occupancy is another source of degeneracy. To see this, consider the Fock states at half-filling with $N_\#$ doubly occupied sites, $N_\#=N_0$ empty sites with $N_\uparrow$ fermions with spin up and $N_\downarrow$ with spin down. There are 
\begin{align}
\begin{pmatrix}
N
\\
N_\#,\, N_0,\, N_\uparrow, N_\downarrow 
\end{pmatrix}
=\frac{N !}{(N_\#!)^2 N_\uparrow ! N_\downarrow !  }
\end{align}
states with this property. This degeneracy also enters into $g$.
Fig.~\ref{fig: heat capacity zero hop} was produced by setting up an efficient algorithm which counts the degeneracies as described.
\section{Fock term contribution}\label{app: Fock term contribution}
\label{sec: app fock term}
We give details of the the mean-field approximation introduced in section \ref{sec: mean-field theory}. Particularly, we demonstrate that it is sufficient to consider the Hartree term whereas the constribution of the Fock term produces a $1/N$ correction. For the Fock term we consider a different contraction of the fermionic operators in the interaction Hamiltonian
\begin{align}
	c^\dagger_{i,\uparrow}c_{i,\downarrow}c^\dagger_{j,\downarrow}c_{j,\uparrow}&\simeq -c^\dagger_{i,\uparrow}c_{j,\uparrow} \langle c^\dagger_{j,\downarrow}c_{i,\downarrow}\rangle-c^\dagger_{j,\downarrow}c_{i,\downarrow}\langle c^\dagger_{i,\uparrow}c_{j,\uparrow}\rangle
	\notag\\
	&+\langle c^\dagger_{i,\uparrow}c_{j,\uparrow}\rangle\langle c^\dagger_{j,\downarrow}c_{i,\downarrow}\rangle,
	\\
	c^\dagger_{i,\downarrow}c_{i,\uparrow}c^\dagger_{j,\uparrow}c_{j,\downarrow} &\simeq -c^\dagger_{i,\downarrow}c_{j,\downarrow}\langle c^\dagger_{j,\uparrow}c_{i,\uparrow}\rangle-c^\dagger_{j,\uparrow}c_{i,\uparrow}\langle c^\dagger_{i,\downarrow}c_{j,\downarrow}\rangle
	\notag\\
	&+\langle c^\dagger_{i,\downarrow}c_{j,\downarrow}\rangle\langle c^\dagger_{j,\uparrow}c_{i,\uparrow}\rangle.
	\label{eq: mean-field_Fock}
\end{align}

Accordingly we can introduce an additional set of order parameters defined as
\begin{align}
	&\Delta_{r} = \langle c^\dagger_{j,\uparrow}c_{j+r,\uparrow}\rangle = \langle c^\dagger_{j,\downarrow}c_{j+r,\downarrow}\rangle,
\end{align}
Here, we made use of spin-rotation symmetry of the original Hamiltonian, which tells us that $\Delta_{r}$ is spin independent. Then, the mean-field Hamiltonian can be written as
\begin{align}
	H_{\mathrm{HF}} = H_t +H_{XY}^{\mathrm{Hartree}}+H_{XY}^{\mathrm{Fock}},
\end{align}
where $H_{\mathrm{MF}} = H_t+H_{XY}^{\mathrm{Hartree}}$ is the mean-field Hamiltonian studied in the main text and the Fock contribution is given by
\begin{align}
H_{XY}^{\mathrm{Fock}} &= \frac{J}{2N}\sum_{i<j,\sigma = \uparrow,\downarrow}\left(c^\dagger_{i\sigma}c_{j,\sigma}+\mathrm{H.c.}\right)\Delta_{|i-j|}
\notag\\
&-\frac{J}{2}\sum_{r=1}^{(N-1)/2}\Delta^2_{r}
\end{align}
We go to momentum space and introduce the $(\pm)$ quasiparticles \eqref{eq: pm quasiparticles}
\begin{align}
	H_{\mathrm{HF}} &= \sum_{k,\pm}\tilde{\varepsilon}_{\pm}(k,\mathcal{M},\Delta_k)c^\dagger_{k,\pm}c_{k,\pm}+\frac{\mathcal{M}^2J}{2}(N-1)\notag
	\\
	&-\frac{J}{2N}\sum_k\Delta_k^2, 
\end{align}

The quasiparticle bands are now given by
\begin{align}
	\tilde{\varepsilon}_{\pm}(k,\mathcal{M},\Delta_k) = -2t\cos(k)\mp\mathcal{M}J\frac{N-1}{2N}+\frac{J\Delta_k}{N},
\end{align} 
Notice, that we have introduced the Fourier transform of the Fock order parameter $\Delta_{k} = \sum_{r}e^{ikr}\Delta_{r}$. At zero temperature the order parameters $\mathcal{M}$ and $\Delta_{k}$ are self-consistently determined minimizing the ground state energy, i.e., imposing the conditions	$\partial E_{\mathrm{HF}}/\partial\mathcal{M} = 0$ and $\partial E_{\mathrm{HF}}/\partial\Delta_k = 0$. The first condition gives Eq.~\eqref{eq:selfconsistent_m} for $\mathcal{M}$, while the second condition tells us that
\begin{align}
	\Delta_{k} = n_{k,+}+n_{k,-} = n_k.
\end{align}
Inserting this result back into the diagonal form of the Hartree-Fock Hamiltonian we obtain
\begin{align}
 	H_{\mathrm{HF}} = H_{\mathrm{MF}}+\frac{J}{2N}\sum_k n_k^2,
\end{align}

where $H_{\mathrm{MF}}$ corresponds to the mean-field Hamiltonian in Eq. ~\eqref{eq:diagonal mean field H}.  
We notice that, the Fock contribution is a finite size correction to the mean-field energy, which can be safely neglected in the thermodynamic limit. In fact the $k$-mode occupation number can take only the values $n_k= 0,1,2$ and then we always have $n_k^2\sim O(1)$. It follows that $\frac{J}{2N}\sum_{k}n_k^2\sim O(1)$. On the other hand $	H_{\mathrm{HF}}$ and $ H_{\mathrm{MF}}$, being extensive quantities, scale as $O(N)$. Accordingly, in the large $N$ limit, we have
\begin{align}
	\frac{H_{\mathrm{HF}}}{N} = \frac{H_{\mathrm{MF}}}{N}+O(N^{-1}).
\end{align}
At finite $N$ the correction due to the Fock contribution adds an energetic penalty to doubly occupied modes with $n_k = 2$ and favors the $XY$ magnetic order. In fact, as shown in the main text, the paramagnetic state has the lowest $N/2$ modes doubly occupied by a $(+)$ type fermion and a $(-)$ type fermion, the paramagnetic Hartree-Fock energy is then
\begin{align}
E_\mathrm{HF}(\mathcal{M}=0) \simeq -\frac{4 Nt}{\pi}+J.
\end{align}
On the contrary, in the ferromagnetic state all the $N$ states are occupied by only one particle, accordingly the ferromagnetic Hartree-Fock energy is
\begin{align}
	E_\mathrm{HF}(\mathcal{M}=1/2) \simeq -\frac{JN}{8}+\frac{J}{2}.
\end{align}
It follows that the correction of order $N^{-1}$ due to the Fock contribution lowers down the value of the critical hopping at finite $N$. 

\end{document}